\documentclass[showpacs,amsmath,superscriptaddress,reprint,aps]{revtex4-1}

\usepackage{graphicx}
\usepackage{hyperref}
\usepackage{color}
\usepackage{float}

\begin{document}

\title
{Universal map of gas-dependent kinetic selectivity in carbon nanotube growth}
\author{K.~Otsuka}
\email[Corresponding author. ]{otsuka@photon.t.u-tokyo.ac.jp}
\affiliation{Department of Mechanical Engineering, The University of Tokyo, Tokyo, 113-8656, Japan}
\affiliation{Nanoscale Quantum Photonics Laboratory, RIKEN Cluster for Pioneering Research, Saitama 351-0198, Japan}
\author{R.~Ishimaru}
\affiliation{Department of Mechanical Engineering, The University of Tokyo, Tokyo, 113-8656, Japan}
\author{A.~Kobayashi}
\affiliation{Department of Mechanical Engineering, The University of Tokyo, Tokyo, 113-8656, Japan}
\author{T.~Inoue}
\affiliation{Department of Applied Physics, Osaka University, Osaka, 565-0871, Japan}
\author{R.~Xiang}
\affiliation{Department of Mechanical Engineering, The University of Tokyo, Tokyo, 113-8656, Japan}
\author{S.~Chiashi}
\affiliation{Department of Mechanical Engineering, The University of Tokyo, Tokyo, 113-8656, Japan}
\author{Y.~K.~Kato}
\affiliation{Nanoscale Quantum Photonics Laboratory, RIKEN Cluster for Pioneering Research, Saitama 351-0198, Japan}
\affiliation{Quantum Optoelectronics Research Team, RIKEN Center for Advanced Photonics, Saitama 351-0198, Japan}
\author{S.~Maruyama}
\email[Corresponding author. ]{maruyama@photon.t.u-tokyo.ac.jp}
\affiliation{Department of Mechanical Engineering, The University of Tokyo, Tokyo, 113-8656, Japan}

\begin{abstract}
Single-walled carbon nanotubes have been a candidate for outperforming silicon in ultrascaled transistors, but the realization of nanotube-based integrated circuits requires dense arrays of purely semiconducting species. Control over kinetics and thermodynamics in tube-catalyst systems plays a key role for direct growth of such nanotube arrays, and further progress requires the comprehensive understanding of seemingly contradictory reports on the growth kinetics. Here, we propose a universal kinetic model and provide its quantitative verification by ethanol-based isotope labeling experiments. While the removal of carbon from catalysts dominates the growth kinetics under a low supply of precursors, our kinetic model and experiments demonstrate that chirality-dependent growth rates emerge when sufficient amounts of carbon and etching agents are co-supplied. As the model can be extended to create kinetic maps as a function of gas compositions, our findings resolve discrepancies in literature and offer rational strategies for chirality selective growth for practical applications.
\end{abstract}

\maketitle
\section{Introduction}
Control over length, density, and chirality of single-walled carbon nanotubes (SWCNTs) at the synthesis stage is still of great interest for their applications. The fabrication of logic integrated circuits (ICs)~\cite{Hills2019,Qiu2018}, for example, requires high-density arrays of exclusively semiconducting species. To fulfill the requirements for digital ICs~\cite{Franklin2013}, significant progress has been made in the chirality sorting in nanotube dispersions and the subsequent assembly~\cite{Brady2016,Liu2020,Jinkins2021}, as well as the selective removal of metallic species~\cite{Jin2013,Shulaker2017,Otsuka2017} from aligned nanotubes grown on crystalline substrates~\cite{Hong2010,Hu2015}. Chirality-sorted nanotubes usually suffer from the excessive interface states that originate from residual surfactants, degrading the switching capability of transistors~\cite{Liu2020,Xu2018}. Ultimately high performance should be achieved by using processing-free nanotube arrays from chemical vapor deposition (CVD) on wafers.

As proposed and experimentally demonstrated~\cite{Ding2009,Artyukhov2014,Yang2014,Zhang2017,Zhu2019,Yang2020,Foerster2021}, the harmony of kinetic and thermodynamic control is crucial to acquire nanotube arrays of a desired electronic type. The abundance of nanotubes with a given chirality ($n$,$m$) is obtained from the product of lengths and population~\cite{Artyukhov2014,Inoue2015}; the understanding of the growth kinetics further involves decoupling of the lengths into growth rates, incubation time, and lifetime (Fig.~\ref{Fig1}a). Several studies have captured the one-by-one growth rates by optical or electron microscopy~\cite{Rao2012,Otsuka2018,He2019,Pimonov2021}, but the results often disagree with each other. Even for the case of alcohols as carbon sources~\cite{Maruyama2002}, the discrepancy is evident. Despite the claim that the growth rate difference is the key for selective growth methods~\cite{Zhou2012,Zhang2017}, one-by-one measurements show that metallic (m-) and semiconducting (s-)SWCNTs exhibit a similar growth rate without a clear chirality dependence, and the growth rate significantly differs within the same chirality~\cite{Otsuka2018,Pimonov2021}.

This confusion may arise from complicated decomposition of precursors~\cite{Xiang2010,Minakov2019}. Since the decomposed products include several carbon sources, as well as etching agents that remove carbon from the catalyst~\cite{Feng2011,Zhou2012}, the growth kinetics cannot be accurately captured without considering these opposing effects. A recent study has shed light on the role of etching by observing the ($n$,$m$)-independent growth rate in the absence of etching agents~\cite{He2019}. There remains vast room for the verification of such a kinetic model and the quantitative understanding of various experimental observations.

Here we propose a universal model that quantitatively describes supply and removal of carbon during a catalytic CVD process, inspired by the distinctive chirality distribution only appearing under a high ethanol vapor pressure. In order to decouple growth parameters usually inter-linked during CVD processes, we develop a duplex labeling technique using isotope ethanol and acetylene, and trace the growth modulation of individual SWCNTs from spatial and spectral points of view. The dominant role of etching agents in the growth kinetics is quantitatively elucidated, which is a key to selective growth according to the electronic structures. The kinetic model also predicts the growth conditions under which the chiral angle-dependence emerges and thereby allows us to experimentally observe the fast growth of near-armchair species. We finally classify the nanotube growth into five regimes depending on the pressures of carbon sources and etching agents, which not only explains various phenomena unresolved in previous studies, but also offers solid strategies for chirality control in catalytic CVD process.

\section{Results and discussion}

\paragraph*{From chirality distribution to growth kinetics of individual nanotubes.}

\begin{figure*}
\includegraphics{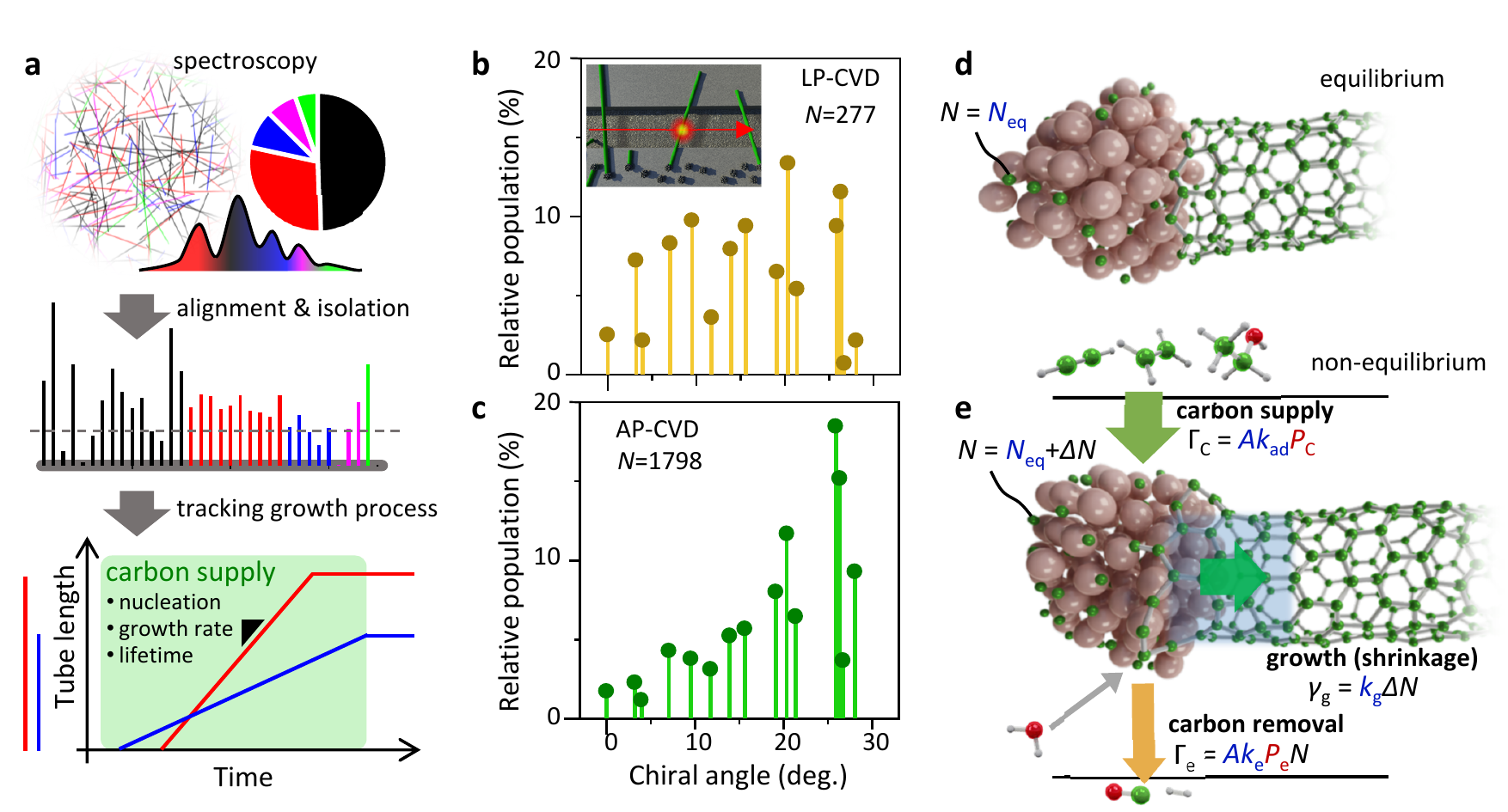}
\caption{
\label{Fig1} From chirality distribution of ensembles to growth kinetics by ``carbon bookkeeping'' of individual nanotubes.
(a) Schematic showing the three levels of analysis on nanotube growth achieved before: spectroscopy to evaluate relative abundance (top), its breakdown into length and population (middle), and the tracing of growth process (bottom). This study seeks to uncover beyond the growth rate. Colors of lines represent nanotube chirality.
(b,c) Relative population of nanotubes with each chirality sorted by chiral angles in the low-pressure ethanol CVD (b) and the atmospheric-pressure ethanol CVD (c), corresponding to the number of SWCNTs whose length exceeds the dashed line in the middle panel of (a). Data in (c) are adopted from ref.~\cite{Otsuka2021}. Inset of (b): Schematic showing the micro-PL measurements of air-suspended SWCNTs over trenches. 
(d,e) Schematics of the kinetic modeling of SWCNT growth at the equilibrium state (d) of a metal nanoparticle and a nanotube wall and that at the non-equilibrium state (e). When no atom enters or leaves the system, carbon concentration $N=N_\mathrm{eq}$. When carbon atoms are supplied to or removed from the iron nanoparticle, carbon concentration $N$ shifts from $N_\mathrm{eq}$, leading to the growth or shrinkage of the nanotube wall. Parameters in dark red is extrinsic and can be determined by experimental setting, while those in dark blue is inherent to the catalyst-nanotube (intrinsic parameters). Sketch of catalyst-nanotube geometries are obtained from our classical molecular dynamics simulation using Tersoff-Brenner potential~\cite{Yoshikawa2019}.
}
\end{figure*}

Before delving into the growth kinetics, we focus on the effect of gas pressure and composition on the chirality distribution of nanotubes grown from ethanol under two slightly different conditions. In both cases, we use evaporated iron as catalysts on thermal oxide of Si substrates and assign the tube chirality by photoluminescence (PL) spectroscopy~\cite{Ishii2015,Otsuka2021}. The key difference lies in the pressure inside the reactor; in the case at a low pressure (LP), ethanol partial pressure is 130 Pa with the flow of Ar containing 3\% of H$_2$. In the other case at an atmospheric pressure (AP), Ar/H$_2$ is supplied through an ethanol bubbler, where ethanol accounts for $\sim$2.4~kPa. While the LP-CVD process yields SWCNTs with a rather uniform distribution (Fig.~\ref{Fig1}b), the clear preference towards near-armchair chirality species has been observed in the AP-CVD (Fig.~\ref{Fig1}c). Because we are observing PL from nanotubes suspended across tranches, chirality-dependent growth rate of nanotubes might lead to the near-armchair preference through the different probability of reaching the other side of trenches. We need to take a closer look at the growth kinetics to understand the modulated chirality distribution.

To analyze the growth kinetics, we consider a steady-state model~\cite{He2019,Puretzky2005} for a nanotube-catalyst system. First, we define the concentration of carbon in a catalyst at an equilibrium state to be $N_\mathrm{eq}$. When no carbon atom is added or removed to/from the catalyst particle, $N$ reaches $N_\mathrm{eq}$ either by precipitation as a tube wall or by dissolution of the tube wall into the catalyst (Fig.~\ref{Fig1}d). In the presence of carbon sources and etching agents, the adsorption and desorption of carbon occur and thereby shift $N$, resulting in the continuous growth or shrinkage of the nanotube (Fig.~\ref{Fig1}e). 

In our model, the growth rate $\gamma_\mathrm{g}$ is proportional to the kinetic constant for growth $k_\mathrm{g}$ and the degree of supersaturation $\Delta N$ ($=N-N_\mathrm{eq}$), 
\begin{equation}
\label{gamma_g}
\gamma_\mathrm{g}=k_\mathrm{g}\Delta N.
\end{equation}
With the focus shifted to the role of etching agents, such as oxygen and water~\cite{Zhang2005,Nasibulin2006,Futaba2009}, carbon atoms are eliminated from the catalyst at the rate $\Gamma_\mathrm{e}$, which is proportional to the surface area $A$, a kinetic constant $k_\mathrm{e}$, the pressure of etching agents $P_\mathrm{e}$, and $N$. The carbon supply rate $\Gamma_\mathrm{C}$ is proportional to $A$, adsorption efficiency $k_\mathrm{ad}$, and the carbon source pressure $P_\mathrm{C}$. We note that a CVD process includes several carbon sources due to gas decomposition, whose $k_\mathrm{ad}$ varies widely, but we do not distinguish those carbon sources and express them using single values for $P_\mathrm{C}$ and $k_\mathrm{ad}$. At the steady state, the equation to describe ``carbon bookkeeping'' becomes,
\begin{equation}
\label{dNdt}
\frac{dN}{dt}\propto \Gamma_\mathrm{C}-(\gamma_\mathrm{g}D+\Gamma_\mathrm{e})=0,
\end{equation}
where $D$ is the number of carbon atoms per unit length ($\propto$ diameter), and therefore the supersaturation of carbon can be expressed as,
\begin{equation}
\label{deltaN}
\Delta N=\frac{k_\mathrm{ad}P_\mathrm{C}-k_\mathrm{e}P_\mathrm{e}N_\mathrm{eq}}{k_\mathrm{g}D'+k_\mathrm{e}P_\mathrm{e}},
\end{equation}
with $D’$ being $D/A$. This kinetic model can describe the growth and shrinkage~\cite{Feng2011,Koyano2019,Pimonov2021} of nanotubes that depend on the balance between $P_\mathrm{C}$ and $P_\mathrm{e}$.

\paragraph*{Dominant role of etching agents in growth kinetics.}

\begin{figure*}
\includegraphics{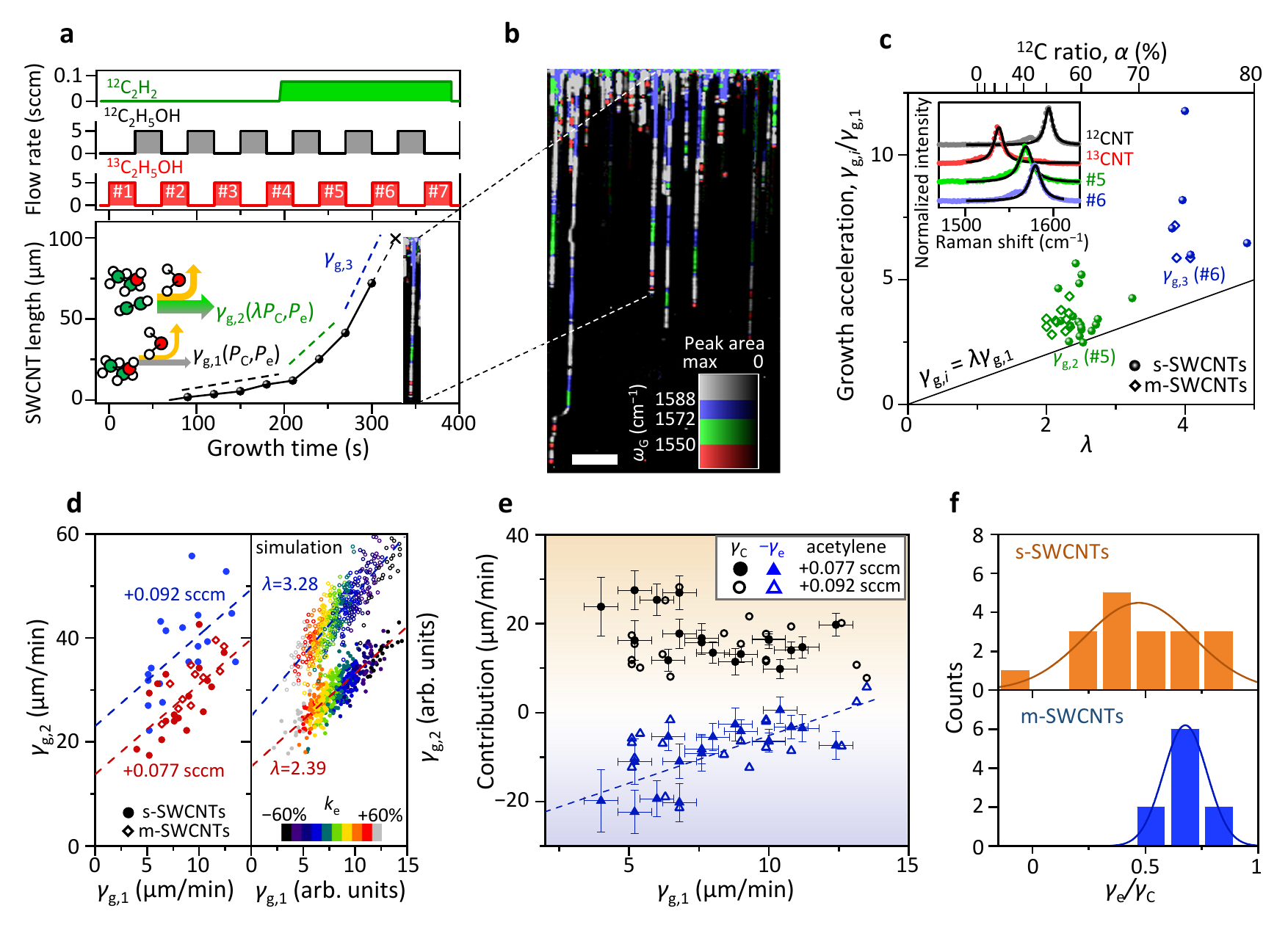}
\caption{
\label{Fig3} ``Carbon bookkeeping'' to elucidate the dominance of carbon removal in low-pressure CVD. 
(a) Switching of feedstock gases during the nanotube growth (upper panel) and a corresponding time evolution of a nanotube length traced by a Raman mapping (lower panel). In addition to Ar/H$_2$ buffer gas at 50~sccm, $^{12}$C ethanol ($^{12}\mathrm{C_2H_5OH}$), $^{13}$C ethanol, and $^{12}$C acetylene ($^{12}\mathrm{C_2H_2}$) are introduced. Acetylene also serves as a second labeling agent that modulates the isotope ratio in nanotubes, allowing identification of the growth time.
(b) Raman mapping image showing the peak position of G-band and its peak area. Catalysts are placed at the top of the image. Excitation wavelength is 532~nm. Scale bar is 20~$\mu$m.
(c) The $P_\mathrm{C}$ multiplication factor $\lambda$ versus the growth rate acceleration $\gamma_{\mathrm{g,}i}/\gamma_\mathrm{g,1}$ ($i=2,3$). The $\lambda$ values are obtained from G-band downshifts that originate from the mixture of $^{12}$C acetylene and $^{13}$C ethanol. Inset: Typical Raman spectra of the nanotube in (a) measured at four different positions. Black lines are the Lorentzian fits.
(d) (Left panel) Experimental growth rates before ($\gamma_\mathrm{g,1}$) and after ($\gamma_\mathrm{g,2}$) the addition of acetylene with two different flow rates (0.077 and 0.092~sccm). (Right panel) Simulated growth rates for comparison at different levels of $\lambda$ (=2.39 and 3.28).
(e) Contribution of carbon adsorption rate $\gamma_\mathrm{C}$ and removal rate $-\gamma_\mathrm{e}$ on the catalyst nanoparticles to the growth rate $\gamma_\mathrm{g,1}$. Data only for s-SWCNTs are plotted.
(f) The ratio of removal rate $\gamma_\mathrm{e}$ to carbon adsorption rate $\gamma_\mathrm{C}$ in the growth of semiconducting (upper) and metallic (lower) nanotubes.
}
\end{figure*}

By varying an extrinsic parameter during the growth and tracing the corresponding modulation of kinetics, we verify the model and determine the intrinsic parameters. Unlike the parameters unique to each catalyst-nanotube pair shown in dark blue in Fig.~\ref{Fig1}e, $P_\mathrm{C}$ and $P_\mathrm{e}$ can be controlled by experimental settings, but they are intricately coupled in the ethanol CVD process. To decouple these parameters, we add a small amount of acetylene gas in the middle of the synthesis, which results in the independent increase of $P_\mathrm{C}$ (Fig.~S3). The complementary experiment where we change only $P_\mathrm{e}$ is summarized in Fig.~S5.

In our duplex labeling technique to trace the growth modulation of individual SWCNTs, ethanol with natural abundance of carbon (hereafter called $^{12}$C ethanol) and $^{13}$C-enriched ethanol (hereafter called $^{13}$C ethanol) are introduced alternately (Fig.~\ref{Fig3}a). We add acetylene of natural abundance (hereafter called $^{12}$C acetylene) as a growth accelerator in the latter half of the CVD process. The lower panel shows a typical time dependence of a nanotube length, which is reconstructed from the Raman mapping (Fig.~\ref{Fig3}b). The growth rate before the acetylene addition ($\gamma_\mathrm{g,1}$) and that defined between the labels \#4 and \#5 ($\gamma_\mathrm{g,2}$) are 5.2 and 29.4~$\mu$m/min, respectively. The growth rate is even accelerated to $\gamma_\mathrm{g,3}=61.2~\mu$m/min after the label \#5, reflecting the slow saturation of a small amount of acetylene in the reactor. The addition of $<$2\% of acetylene to ethanol accelerates the growth by 12-fold owing to its efficient adsorption on the catalyst~\cite{Xiang2009}. 

To fully exploit the model, we should quantitatively determine the multiplication factor $\lambda$ of the total supply rate of carbon, assuming the acetylene addition increases $P_\mathrm{C}$ to $\lambda P_\mathrm{C}$. The isotope ratio in the nanotube grown from the mixture of $^{13}$C ethanol and $^{12}$C acetylene provides accurate $\lambda$ for each nanotube through Raman spectroscopy (the inset of Fig.~\ref{Fig3}c). The $^{12}$C ratio $\alpha$ in the label \#5 of the particular nanotube (Fig.~\ref{Fig3}a) is calculated to be 59\% from the G-band frequency. The $P_\mathrm{C}$ multiplication factor is then $\lambda=2.42$ ($=1/(1-\alpha)$). To understand how the increased $P_\mathrm{C}$ affects the actual growth rate, we plot the growth acceleration factor $\gamma_{\mathrm{g,}i}/\gamma_\mathrm{g,1}$ ($i=2,3$) against $\lambda$ for all nanotubes (Fig.~\ref{Fig3}c). It is somewhat surprising that the growth rates after the acetylene addition is noticeably larger than $\lambda \gamma_\mathrm{g,1}$ (solid line). Carbon removal from the catalyst holds the key to this discrepancy because the simple rate equations are derived for the growth before and after the acetylene addition,
\begin{equation}
\label{simple-g1}
\gamma_\mathrm{g,1}=\gamma_\mathrm{C}-\gamma_\mathrm{e},
\end{equation}
\begin{equation}
\label{simple-g2}
\gamma_\mathrm{g,2}=\lambda \gamma_\mathrm{C}-s\gamma_\mathrm{e},
\end{equation}
where $\gamma_\mathrm{C}$ and $\gamma_\mathrm{e}$ are defined as $\Gamma_\mathrm{C}/D$ and $\Gamma_\mathrm{e}/D$, respectively. The $\gamma_\mathrm{e}$ multiplication factor due to increased $P_\mathrm{C}$ is given by $s=N_2/N_1=\frac{\lambda k_\mathrm{ad}P_\mathrm{C}+k_\mathrm{g}D'N_\mathrm{eq}}{k_\mathrm{ad}P_\mathrm{C}+k_\mathrm{g}D'N_\mathrm{eq}}$ ($1<s<\lambda$)  with $N_1$ ($N_2$) being $N$ before (after) the acetylene addition.

The left panel of Fig.~\ref{Fig3}d shows the accelerated growth rate $\gamma_\mathrm{g,2}$ plotted against $\gamma_\mathrm{g,1}$ with two different flow rates of additive acetylene. Interestingly, the linear fit of each set of experimental data has a large value of y-intercept. For better understating of this tendency, we derive analytical expressions of $\gamma_\mathrm{g,1}$ and $\gamma_\mathrm{g,2}$ by assigning $P_\mathrm{C}$ and $\lambda P_\mathrm{C}$ to $P_\mathrm{C}$ in Eq.~\ref{deltaN}, respectively, and then remove $k_\mathrm{e}$ to yield the relationship, 
\begin{equation}
\label{g2-g1}
\gamma_\mathrm{g,2}=\frac{\lambda k_\mathrm{ad}P_\mathrm{C}+k_\mathrm{g}D'N_\mathrm{eq}}{k_\mathrm{ad}P_\mathrm{C}+k_\mathrm{g}D'N_\mathrm{eq}}\gamma_\mathrm{g,1}+\frac{k_\mathrm{ad}P_\mathrm{C}k_\mathrm{g}N_\mathrm{eq}}{k_\mathrm{ad}P_\mathrm{C}+k_\mathrm{g}D'N_\mathrm{eq}}(\lambda-1),
\end{equation}
which assumes $k_\mathrm{e}$ variance to account for the $\gamma_\mathrm{g}$ difference. The experimental confirmation of large y-intercepts, which correspond to the second term on the right-hand side of Eq.~\ref{g2-g1}, is equivalent to a non-zero $N_\mathrm{eq}$. This supports the versatility of our kinetic model to include nanotube shrinkage as well as growth. Note that eliminating $k_\mathrm{e}$ from the $\gamma_\mathrm{g,1}$-$\gamma_\mathrm{g,2}$ relationship yields the best fit to the experimental results (see Fig.~S9). With two different amounts of additive acetylene, the ratio of the y-intercepts (14.4/23.0) agrees to that independently obtained from the average $\lambda-1$ values (1.39/2.28), also verifying our kinetic model. 

Having obtained $\lambda$ from Raman spectra, we reproduce the $\gamma_\mathrm{g,1}$-$\gamma_\mathrm{g,2}$ relationship using our kinetic model. When we simulate  $\gamma_\mathrm{g}$ with the carbon source pressures of $P_\mathrm{C}$ and $\lambda P_\mathrm{C}$, the relationships in the right panel of Fig.~\ref{Fig3}d are obtained. Here, $k_\mathrm{e}$ and $k_\mathrm{g}$ are determined from the averages of $\gamma_\mathrm{g,1}$ and $\gamma_\mathrm{g,2}$, and we assume both kinetic constants have relative standard deviations of 20\%. By comparing the experiments and simulations, the variance in $\gamma_\mathrm{g,1}$ can be attributed to $k_\mathrm{e}$ varying from the average by up to $\pm$60\%. Although the parameters are adjusted using the results with the flow rate of 0.077~sccm alone, we can reasonably predict another $\gamma_\mathrm{g,1}$-$\gamma_\mathrm{g,2}$ distribution for the other amount of acetylene once $\lambda$ is determined from the Raman spectra.

\begin{figure*}
\includegraphics{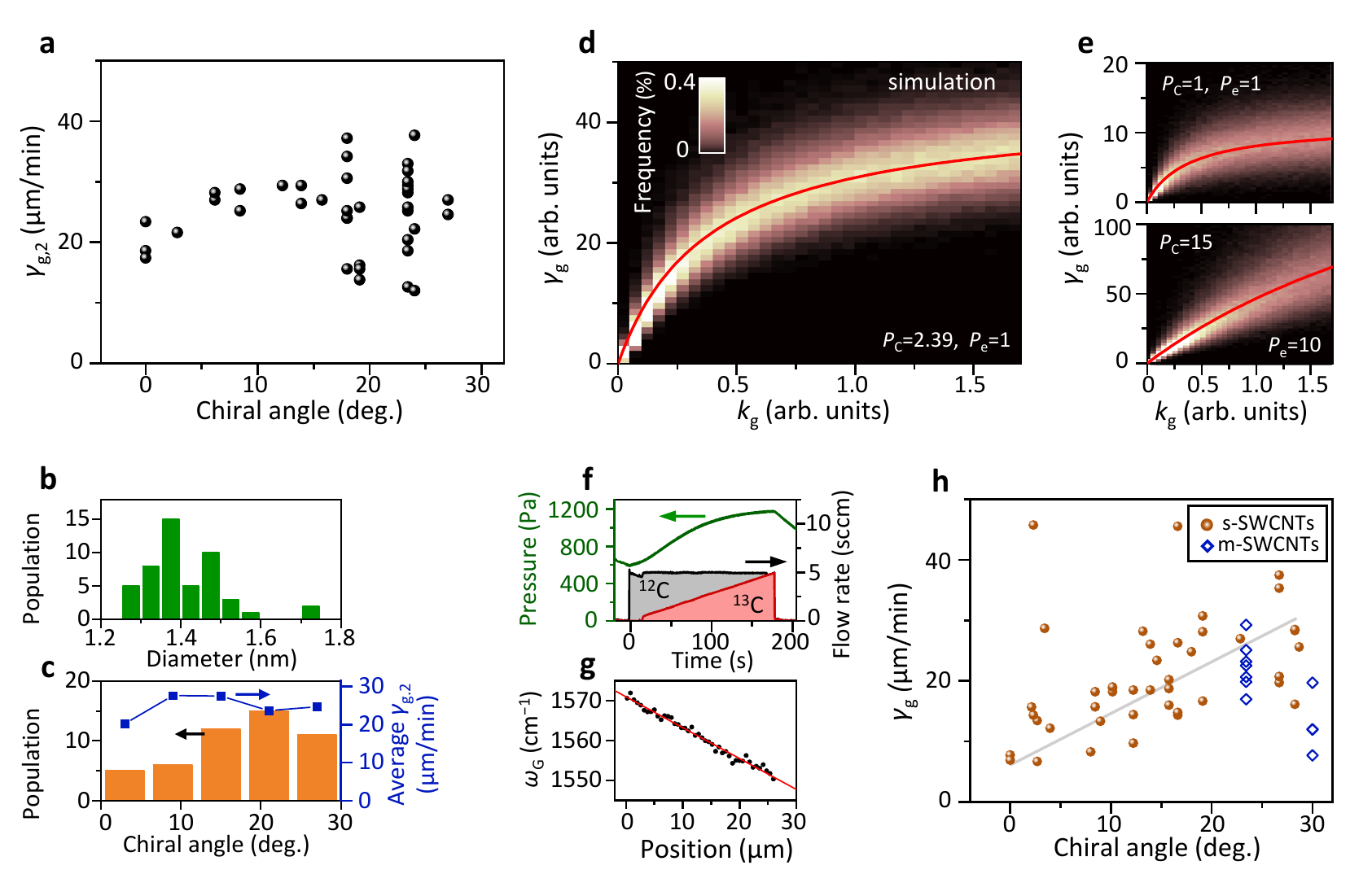}
\caption{
\label{Fig4} Absence and emergence of chiral angle-dependent growth kinetics.
(a) Growth rate $\gamma_\mathrm{g,2}$ right after the acetylene addition as a function of the chiral angle.
(b,c) Diameter (b) and chiral angle (c) distribution of analyzed SWCNTs.
(d) Deterministic growth rate (red solid line) and simulated growth rate distribution (color contour map) as a function of the kinetic constant $k_\mathrm{g}$ of the catalyst-nanotube systems. The parameters that reproduce the growth rate distribution in Fig.~\ref{Fig3}d ($\lambda=2.39$) are used with $k_\mathrm{g}$ uniformly distributed between 0 and 1.7, where $k_\mathrm{g}=0.85$ corresponds to the average in the experiments.
(e) Simulated growth rate distribution with the conditions that emulate the purely ethanol-based CVD (upper panel) and with $15\times$ carbon sources and $10\times$ etching agents (lower panel).
(f) Total pressure and ethanol flow rate during the isotope labeling CVD, where $^{13}$C ratio monotonically increases with time. As the total flow rate is small, extended dwell time of ethanol in the furnace leads to an increased $P_\mathrm{C}/P_\mathrm{e}$ ratio due to acetylene generation (Fig.~S3).
(g) Typical G-band peak position $\omega_\mathrm{G}$ plotted against the position along the tube axis.
(h) Growth rate of the nanotube grown with the isotope labeling as a function of the chiral angle. Orange circles and blue diamonds represent s- and m-SWCNTs, respectively. The gray line is a guide for the eye.
}
\end{figure*}

We take a closer look at the influence of $\Gamma_\mathrm{e}$ in individual nanotubes.  As the slope of the $\gamma_\mathrm{g,1}$-$\gamma_\mathrm{g,2}$ relationship is equal to $s$ (Eq.~\ref{g2-g1}), we can breakdown the growth rate $\gamma_\mathrm{g,1}$ of each nanotube into the contribution of supply and removal at a catalyst using Eq.~\ref{simple-g1} and \ref{simple-g2} (Fig.~\ref{Fig3}e). The growth rate is correlated with the rate of carbon removal from the catalyst, indicating the dominant influence of the etching agent derived from ethanol on growth rate determination under this growth condition.

Such carbon removal effects have been claimed to be the driving force behind the selective growth of s-SWCNTs~\cite{Ding2009a,Zhou2012}. Using the duplex labeling, we can successfully quantify the susceptivity to carbon removal in the form of $\gamma_\mathrm{e}/\gamma_\mathrm{C}$ for s- and m-SWCNTs as shown in Fig.~\ref{Fig3}f. While s-SWCNTs have a wide distribution between 0 to 1, m-SWCNTs show a rather narrow distribution closer to 1. We expect that when $P_\mathrm{e}$ is further increased, the nanotubes with a large $\gamma_\mathrm{e}/\gamma_\mathrm{C}$, i.e., most of m-SWCNTs and a part of s-SWCNTs, start to be shortened, leading to the preferential growth of semiconducting species. 

\paragraph*{Emergence of chirality-dependent growth kinetics.}

So far, we have discussed the carbon supply/removal for catalysts. One might wonder if the susceptivity to carbon removal is determined by chirality ($n,m$)~\cite{Kimura2018}, but that is excluded by the widely distributed $\gamma_\mathrm{e}/\gamma_\mathrm{C}$ (0.15--0.68) within the (12,8) nanotubes (Fig.~S11). We can confirm the chirality-independent $\Gamma_\mathrm{e}$ from the growth rate $\gamma_\mathrm{g,2}$ after the acetylene addition, which is plotted against the chiral angle (Fig.~\ref{Fig4}a). The dispersion within similar chiral angles is large likely due to the non-uniformity of catalyst particles, and no overall trend can be found. Figure~\ref{Fig4}b and c show the diameter and chiral angle distribution. Judging from the more significant preference towards armchair chirality in terms of population than that in the growth kinetics, the weak bias in the chiral-angle distribution in Fig.~\ref{Fig1}b should be attributed to the number of nucleation in this regime. 

With the kinetic model, we can clearly understand the insignificance of ($n$,$m$) in the apparent growth rate. In Fig.~\ref{Fig4}d, we draw the deterministic growth rate 
\begin{equation}
\label{detail_gamma}
\gamma_\mathrm{g}=k_\mathrm{g}\Delta N=\frac{k_\mathrm{g}(k_\mathrm{ad}P_\mathrm{C}-k_\mathrm{e}P_\mathrm{e}N_\mathrm{eq})}{k_\mathrm{g}D'+k_\mathrm{e}P_\mathrm{e}}.
\end{equation}
From the simulation in which $k_\mathrm{e}$ follows a normal distribution with a deviation of 20\%, the distribution of growth rate at a certain $k_\mathrm{g}$ can be obtained and overlayed in Fig.~\ref{Fig4}d as a color contour. The growth rate easily saturates with respect to $k_\mathrm{g}$ because carbon supply to the catalyst becomes the rate-determining step at a small $P_\mathrm{e}$. Furthermore, $\gamma_\mathrm{g}$ varies widely depending on $k_\mathrm{e}$ (at a fixed $k_\mathrm{g}$), indicating that $k_\mathrm{g}$ is not the dominant factor in determining the growth kinetics. The absence of an explicit trend in the experiment (Fig.~\ref{Fig4}a) can be reasonably explained by these two aspects. We can also simulate the  $k_\mathrm{g}$-dependent growth rate for the condition corresponding to the ethanol CVD that yields $\gamma_\mathrm{g,1}$ (the upper panel of Fig.~\ref{Fig4}e). At the reduced $P_\mathrm{C}/P_\mathrm{e}$ ratio, the relative dispersion at a fixed $k_\mathrm{g}$ is even larger, in good agreement with our previous study~\cite{Otsuka2018}. 

According to the above comparison, the clear ($n$,$m$) dependence in growth rates should appear when both $P_\mathrm{e}$ and $P_\mathrm{C}/P_\mathrm{e}$ are increased, and the corresponding distribution of simulated growth rates is shown in the lower panel of Fig.~\ref{Fig4}e. Experimentally, such a growth condition is often seen in the ethanol-based AP-CVD, where a large amount of acetylene is generated through a long-time dwelling in the hot zone of furnace (Fig.~S2 and S3). 

To test the above hypothesis, we conduct another isotope labeling experiment, where we emulate the conditions for the AP-CVD~\cite{Ishii2015} by introducing ethanol without a carrier gas (Fig.~\ref{Fig4}f). As the total pressure is similar to the CVD process for Fig.~\ref{Fig3} despite the absence of buffer gases, the partial pressure of ethanol and the gas heating time are $10\times$ larger. In this experiment, after reducing the catalyst in Ar/H$_2$ atmosphere, we start the synthesis only with $^{12}$C ethanol and gradually increase the fraction of $^{13}$C ethanol. Figure~\ref{Fig4}g shows the peak frequency of G-band obtained at different positions along a nanotube, slope of which can be converted into a growth rate. As shown in Fig.~\ref{Fig4}h, the growth rate is more clearly correlated to the chiral angle, though armchair ($n$,$n$) nanotubes~\cite{Artyukhov2014} and a few s-SWCNTs deviate from the overall trend (see Fig.~S2). Compared with the result in Fig.~\ref{Fig4}a, the growth rate with a small chiral angle is relatively suppressed in this growth regime. 

\paragraph*{Kinetic regimes depending on gas compositions.}

\begin{figure*}
\includegraphics[scale=0.95]{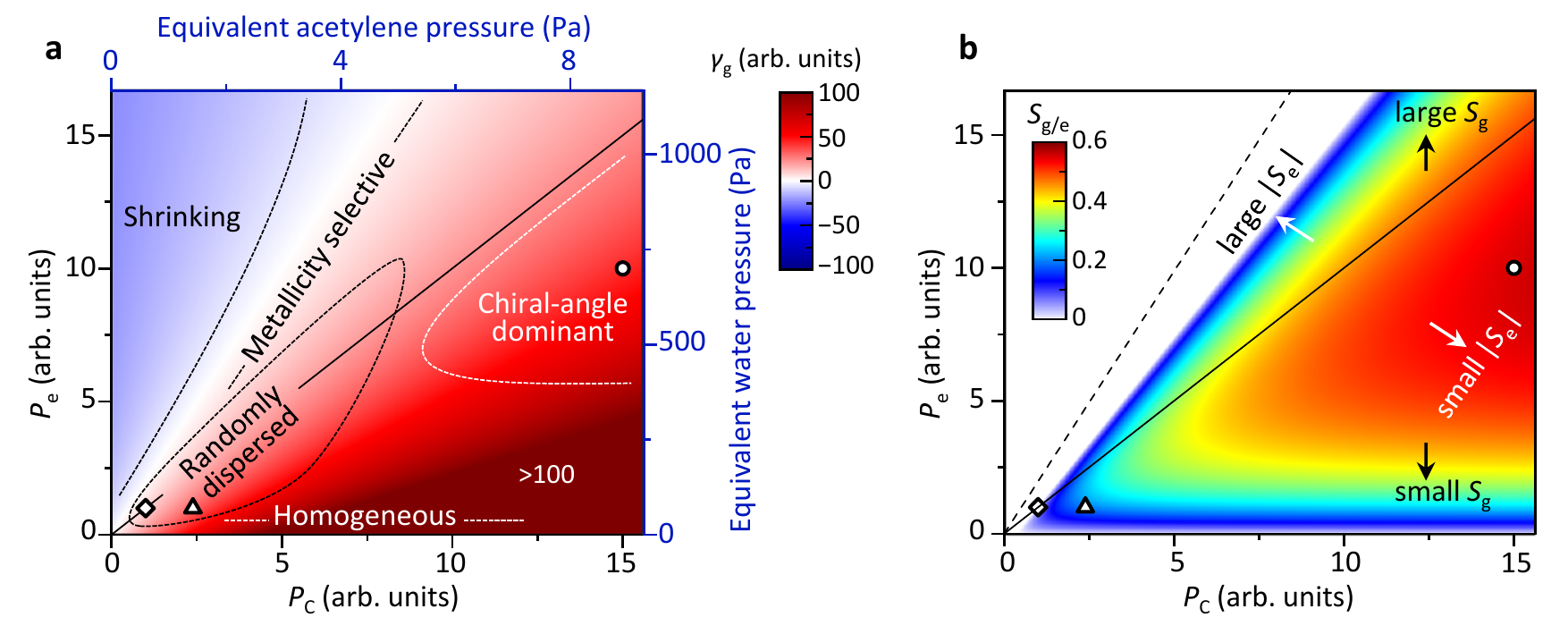}
\caption{
\label{Fig5} Growth kinetic regimes that depend on pressures of carbon sources and etching agents.
(a) Model-based growth rate as a function of the carbon source pressure $P_\mathrm{C}$ and etching agent pressure $P_\mathrm{e}$. Equivalent pressures of acetylene and water are based on $P_\mathrm{C}$ increase by added acetylene and the decomposition simulation of ethanol (Fig.~S3), respectively.
(b) Dominance of the kinetic constant for growth $S_\mathrm{g/e}$ in determining growth rate as a function of $P_\mathrm{C}$ and $P_\mathrm{e}$. Diamond, triangle, and circle marks represent the growth conditions that yield $\gamma_\mathrm{g,1}$ and $\gamma_\mathrm{g,2}$ in Fig.~\ref{Fig3}a, and $\gamma_\mathrm{g}$ in Fig.~\ref{Fig4}h, respectively. Dashed line and solid line represent the balancing point with zero growth rates and the same $P_\mathrm{C}/P_\mathrm{e}$ ratio as in the ethanol-based LP-CVD ($\gamma_\mathrm{g,1}$), respectively.
}
\end{figure*}

By taking gas compositions as explanatory variables, we can generalize the above discussions to gain a full picture of growth kinetics. Figure~\ref{Fig5}a shows the growth rate at the fixed $k_\mathrm{g}$ and $k_\mathrm{e}$ as a function of $P_\mathrm{C}$ and $P_\mathrm{e}$. In addition, the sensitivities $S_\mathrm{g}$ ($S_\mathrm{e}$) of $\gamma_\mathrm{g}$ to the change in $k_\mathrm{g}$ ($k_\mathrm{e}$) are obtained from the equations $S_g=\frac{\partial \gamma_\mathrm{g}}{\partial k_\mathrm{g}} \frac{k_\mathrm{g}}{|\gamma_\mathrm{g}|}$ ($S_\mathrm{e}=\frac{\partial \gamma_\mathrm{g}}{\partial k_\mathrm{e}} \frac{k_\mathrm{e}}{|\gamma_\mathrm{g}|}$). Similar two-dimensional maps for $S_\mathrm{g}$ and $S_\mathrm{e}$ are provided in Fig.~S12. Using these sensitivities, we define the relative dominance of $k_\mathrm{g}$ in determining growth rates by $S_\mathrm{g/e}=S_\mathrm{g}+\zeta S_\mathrm{e}$ (a weight coefficient $\zeta$ of 0.2 is chosen) as shown in Fig.~\ref{Fig5}b. A large $S_\mathrm{g/e}$ requires a large $P_\mathrm{C}/P_\mathrm{e}$ ratio and a large $P_\mathrm{e}$ to suppress $S_\mathrm{e}$ while keeping $S_\mathrm{g}$ large.

When hydrocarbon is supplied and $P_\mathrm{e}$ is negligible, the growth rate is simply limited by the carbon supply rate. All nanotubes thus grow at a similar rate as seen in the previous studies (``homogeneous rate regime'')~\cite{Wang2018,Hussain2018,He2018}. In this regime, a smaller $k_\mathrm{g}$ results in a short growth lifetime due to a large $N$, and the ($n$,$m$) dependence may appear rather in the lifetime difference~\cite{He2019}. In contrast, ``randomly dispersed rate regime'' should be observed in the low-pressure ethanol CVD, where $P_\mathrm{C}$ and $P_\mathrm{e}$ are small. Growth rates are largely influenced by $k_\mathrm{e}$, an uncontrolled property of catalyst-nanotube pairs, leading to a minor ($n$,$m$) dependence. Unlike the homogeneous rate regime, the catalyst with a large $k_\mathrm{e}$ is expected to have a small $N$ and hence a long lifetime, which can explain the negative correlation between $\gamma_\mathrm{g}$ and lifetime~\cite{Pimonov2021} (Fig.~S13). Historically, the above two regimes with a small $P_\mathrm{e}$ appeared when one attempted to capture the suppressed growth rate under somewhat extreme conditions (e.g., in-situ electron microscopy)~\cite{Takagi2006,He2019}. This may ironically led to the disappearance of the chiral angle-dependent growth rate particularly in the experiments that carefully studied individual SWCNTs~\cite{Inoue2015,Otsuka2018,Pimonov2021}. 

When the partial pressure of ethanol increases, $P_\mathrm{C}$ and $P_\mathrm{e}$ accordingly increases with their ratio being constant as drawn by the solid line in Fig.~\ref{Fig5}b. We have previously observed that increasing ethanol pressure alone leads to a significant tube-to-tube dispersion in the growth rate~\cite{Inoue2013,Otsuka2018} and saturation of its average value~\cite{Einarsson2008} (Fig.~S14). Although $P_\mathrm{C}$ and $P_\mathrm{e}$ are inter-related when using only ethanol, independent control of $P_\mathrm{e}$ has been achieved by adding water vapor or methanol~\cite{Ding2009a,Zhou2012}. With the addition of such etching agents, the reduced $P_\mathrm{C}/P_\mathrm{e}$ ratio results in the ``metallicity selective regime'', where only the nanotubes with a small $k_\mathrm{e}$ can grow longer. As shown in Fig.~\ref{Fig3}f, m-SWCNTs have a relatively high $k_\mathrm{e}$, and therefore the growth of metallic nanotubes is preferentially suppressed. When the $P_\mathrm{C}/P_\mathrm{e}$ ratio becomes even smaller, all nanotubes will be shortened (``shrinking regime'') as experimentally observed in the literature~\cite{Feng2011}. It is noteworthy that growth and shrinkage rates have different dependencies on $k_\mathrm{g}$ and $k_\mathrm{e}$~\cite{Koyano2019} (Fig.~S13).

Finally, the chiral-angle dominant rate regime~\cite{Rao2012} emerges when both the $P_\mathrm{C}/P_\mathrm{e}$ ratio and $P_\mathrm{e}$ are large. The white circles in Fig.~\ref{Fig5} corresponding to the growth condition for Fig.~\ref{Fig4}f--h falls under this regime. Note that when CO is used as a precursor, the reverse reaction of disproportionation easily happens, resulting in the preferred growth of near-armchair species from liquid catalysts without explicitly adding etching agents~\cite{He2019,Bachilo2002,Liao2018}. As the previous studies that reported the growth of ($n$,$m$)-specific nanotubes employed the ethanol bubbler-based AP-CVD~\cite{Yang2014,Zhang2017}, we suspect that the growth rates highest in a ($2n$,$n$) chirality indeed played an important role in solid catalyst systems, in addition to a thermodynamic preference. 

In conclusion, we have proposed the simple but universal kinetic model of nanotube growth and provided the experimental supports. Whereas the susceptivity to carbon removal governs the growth rates in the low-pressure ethanol CVD, the kinetic model predicts the emergence of the chirality-dependent growth rates with the increased ethanol pressure. Furthermore, by generalizing the catalytic growth from the viewpoints of carbon supply and removal, we are able to comprehensively explain various experimental results that previously seemed contradictory. Such kinetic maps could be extended for a wide range of temperature and catalyst systems. For the selective growth with the purity level required for logic IC applications, the dependence of $k_\mathrm{ad}$ and $k_\mathrm{e}$ on the electronic structure of catalyst-nanotube systems also needs to be studied to boost kinetic selectivity. To capture transient phenomena at an early growth stage, atomic-scale experiments and theoretical study should be performed, which will provide rational strategies for the control over SWCNT arrays when combined with the present findings.

\section*{Methods}
\paragraph*{Carbon nanotube growth.}
SWCNT arrays are grown on r-cut single-crystal quartz substrates (Hoffman Materials Inc.)~\cite{Otsuka2018}, and air-suspended SWCNTs are grown across trenches that are formed on SiO$_2$/Si substrates by dry etching~\cite{Otsuka2021}.  Iron catalysts with a typical thickness of 0.1~nm are evaporated in lithographically patterned areas. The catalyst is reduced in an Ar atmosphere containing 3\% H$_2$ at 800$^{\circ}$C for 10~min, followed by the supply of carbon sources. Total pressure during the growth is 1.3--1.5~kPa and 110~kPa for the LP- and AP-CVD process, respectively. The flow rates of Ar/H$_2$ are 50 and 500~sccm for isotope labeling experiments and normal growth experiments without labeling, respectively. In the AP-CVD process, ethanol is supplied from a bubbler kept at 5$^\circ$C using Ar/H$_2$ as a carrier gas. After a certain growth duration, the carbon source supply is stopped, and the furnace is cooled to room temperature. Ethanol and acetylene with the natural isotope ratio ($^{12}$C $\sim$99\%), as well as $^{13}$C-enriched ethanol (Cambridge Isotope Laboratories, Inc., 1,2-$^{13}$C$_2$, 99\%) are used as carbon sources. Details for each growth condition are summarized in Table~S1. 

\paragraph*{Raman mapping and spectroscopy of aligned nanotubes.}
Raman spectroscopy is performed for nanotube samples transferred to SiO$_2$/Si substrates. SWCNT arrays are transferred via poly(methyl methacrylate) to Si with a 100-nm-thick oxide layer. To locate SWCNTs, metallic markers (Ti and Pt) are patterned on SiO$_2$/Si substrates before the transfer of the SWCNTs. We use Raman spectrometry (Renishaw, inVia) to determine the types and positions of isotope labels in SWCNTs transferred to SiO$_2$/Si substrates and then converted them to time evolution of SWCNT lengths~\cite{Otsuka2018}. Raman spectra are obtained in 0.6~$\mu$m steps along directions both parallel and perpendicular to the SWCNT orientation. Excitation wavelengths of 488 and 532~nm are mainly used. This is because the strong power of the available laser enabled efficient Raman mapping measurement, and the photon energies are resonant with SWCNTs of various chirality grown under the current growth conditions. Typically, each Raman spectrum is acquired over 5~s with excitation by a $\sim0.6~\mu$m wide and $\sim17~\mu$m long laser spot with an intensity of $\sim30~$mW (power density $\sim3\times10^{5}~\mathrm{W}/\mathrm{cm}^2$).

\paragraph*{PL measurement of air-suspended nanotubes.}
A homebuilt confocal microscopy system is used to perform PL measurements at room temperature in air~\cite{Ishii2015,Otsuka2021}. We use a wavelength-tunable Ti:sapphire laser for excitation. The laser beam is focused on the samples using an objective lens with a numerical aperture of 0.65, and a working distance of 4.5~mm. PL is collected through the same objective lens and detected using a liquid-nitrogen-cooled InGaAs diode array attached to a spectrometer. For the quick chirality assignment, excitation wavelengths of 780, 850, and 910~nm are used, which are nearly resonant to a wide range of nanotubes with diameters between 0.98--1.36~nm. Excitation powers of 10--20~$\mu$W are used (see Fig.~S1).

\begin{acknowledgments}
Part of this work was supported by JSPS (KAKENHI JP20K15137, JP20H00220) and JST (CREST JPMJCR20B5). A part of this work was conducted at Takeda Sentanchi Supercleanroom, The University of Tokyo, supported by "Nanotechnology Platform Program" of the Ministry of Education, Culture, Sports, Science and Technology (MEXT), Japan, Grant Number JPMXP09F21UT0047.
\end{acknowledgments}

\section*{Author Contributions}
K.O. and S.M. conceived the project and designed the experiments. K.O., R.I., and A.K. synthesized the samples and carried out the Raman spectroscopy measurements. K.O. and Y.K.K. performed photoluminescence spectroscopy. K.O., R.I., and A.K. analyzed data. K.O. proposed the growth kinetic model and wrote the manuscript. T.I., R.X., S.C., Y.K.K., and S.M. commented on the manuscript.


\begin{thebibliography}{48}%
\makeatletter
\providecommand \@ifxundefined [1]{%
 \@ifx{#1\undefined}
}%
\providecommand \@ifnum [1]{%
 \ifnum #1\expandafter \@firstoftwo
 \else \expandafter \@secondoftwo
 \fi
}%
\providecommand \@ifx [1]{%
 \ifx #1\expandafter \@firstoftwo
 \else \expandafter \@secondoftwo
 \fi
}%
\providecommand \natexlab [1]{#1}%
\providecommand \enquote  [1]{``#1''}%
\providecommand \bibnamefont  [1]{#1}%
\providecommand \bibfnamefont [1]{#1}%
\providecommand \citenamefont [1]{#1}%
\providecommand \href@noop [0]{\@secondoftwo}%
\providecommand \href [0]{\begingroup \@sanitize@url \@href}%
\providecommand \@href[1]{\@@startlink{#1}\@@href}%
\providecommand \@@href[1]{\endgroup#1\@@endlink}%
\providecommand \@sanitize@url [0]{\catcode `\\12\catcode `\$12\catcode
  `\&12\catcode `\#12\catcode `\^12\catcode `\_12\catcode `\%12\relax}%
\providecommand \@@startlink[1]{}%
\providecommand \@@endlink[0]{}%
\providecommand \url  [0]{\begingroup\@sanitize@url \@url }%
\providecommand \@url [1]{\endgroup\@href {#1}{\urlprefix }}%
\providecommand \urlprefix  [0]{URL }%
\providecommand \Eprint [0]{\href }%
\providecommand \doibase [0]{http://dx.doi.org/}%
\providecommand \selectlanguage [0]{\@gobble}%
\providecommand \bibinfo  [0]{\@secondoftwo}%
\providecommand \bibfield  [0]{\@secondoftwo}%
\providecommand \translation [1]{[#1]}%
\providecommand \BibitemOpen [0]{}%
\providecommand \bibitemStop [0]{}%
\providecommand \bibitemNoStop [0]{.\EOS\space}%
\providecommand \EOS [0]{\spacefactor3000\relax}%
\providecommand \BibitemShut  [1]{\csname bibitem#1\endcsname}%
\let\auto@bib@innerbib\@empty
%</preamble>
\bibitem [{\citenamefont {Hills}\ \emph {et~al.}(2019)\citenamefont {Hills},
  \citenamefont {Lau}, \citenamefont {Wright}, \citenamefont {Fuller},
  \citenamefont {Bishop}, \citenamefont {Srimani}, \citenamefont {Kanhaiya},
  \citenamefont {Ho}, \citenamefont {Amer}, \citenamefont {Stein},
  \citenamefont {Murphy}, \citenamefont {Arvind}, \citenamefont
  {Chandrakasan},\ and\ \citenamefont {Shulaker}}]{Hills2019}%
  \BibitemOpen
  \bibfield  {author} {\bibinfo {author} {\bibfnamefont {G.}~\bibnamefont
  {Hills}}, \bibinfo {author} {\bibfnamefont {C.}~\bibnamefont {Lau}}, \bibinfo
  {author} {\bibfnamefont {A.}~\bibnamefont {Wright}}, \bibinfo {author}
  {\bibfnamefont {S.}~\bibnamefont {Fuller}}, \bibinfo {author} {\bibfnamefont
  {M.~D.}\ \bibnamefont {Bishop}}, \bibinfo {author} {\bibfnamefont
  {T.}~\bibnamefont {Srimani}}, \bibinfo {author} {\bibfnamefont
  {P.}~\bibnamefont {Kanhaiya}}, \bibinfo {author} {\bibfnamefont
  {R.}~\bibnamefont {Ho}}, \bibinfo {author} {\bibfnamefont {A.}~\bibnamefont
  {Amer}}, \bibinfo {author} {\bibfnamefont {Y.}~\bibnamefont {Stein}},
  \bibinfo {author} {\bibfnamefont {D.}~\bibnamefont {Murphy}}, \bibinfo
  {author} {\bibnamefont {Arvind}}, \bibinfo {author} {\bibfnamefont
  {A.}~\bibnamefont {Chandrakasan}}, \ and\ \bibinfo {author} {\bibfnamefont
  {M.~M.}\ \bibnamefont {Shulaker}},\ }\bibfield  {title} {\bibinfo {title}
  {Modern microprocessor built from complementary carbon nanotube
  transistors},\ }\href {\doibase 10.1038/s41586-019-1493-8} {\bibfield
  {journal} {\bibinfo  {journal} {Nature}\ }\textbf {\bibinfo {volume} {572}},\
  \bibinfo {pages} {595} (\bibinfo {year} {2019})}\BibitemShut {NoStop}%
\bibitem [{\citenamefont {Qiu}\ \emph {et~al.}(2018)\citenamefont {Qiu},
  \citenamefont {Liu}, \citenamefont {Xu}, \citenamefont {Deng}, \citenamefont
  {Xiao}, \citenamefont {Si}, \citenamefont {Lin}, \citenamefont {Zhang},
  \citenamefont {Wang}, \citenamefont {Guo}, \citenamefont {Peng},\ and\
  \citenamefont {Peng}}]{Qiu2018}%
  \BibitemOpen
  \bibfield  {author} {\bibinfo {author} {\bibfnamefont {C.}~\bibnamefont
  {Qiu}}, \bibinfo {author} {\bibfnamefont {F.}~\bibnamefont {Liu}}, \bibinfo
  {author} {\bibfnamefont {L.}~\bibnamefont {Xu}}, \bibinfo {author}
  {\bibfnamefont {B.}~\bibnamefont {Deng}}, \bibinfo {author} {\bibfnamefont
  {M.}~\bibnamefont {Xiao}}, \bibinfo {author} {\bibfnamefont {J.}~\bibnamefont
  {Si}}, \bibinfo {author} {\bibfnamefont {L.}~\bibnamefont {Lin}}, \bibinfo
  {author} {\bibfnamefont {Z.}~\bibnamefont {Zhang}}, \bibinfo {author}
  {\bibfnamefont {J.}~\bibnamefont {Wang}}, \bibinfo {author} {\bibfnamefont
  {H.}~\bibnamefont {Guo}}, \bibinfo {author} {\bibfnamefont {H.}~\bibnamefont
  {Peng}}, \ and\ \bibinfo {author} {\bibfnamefont {L.-M.}\ \bibnamefont
  {Peng}},\ }\bibfield  {title} {\bibinfo {title} {Dirac-source field-effect
  transistors as energy-efficient, high-performance electronic switches},\
  }\href {\doibase 10.1126/science.aap9195} {\bibfield  {journal} {\bibinfo
  {journal} {Science}\ }\textbf {\bibinfo {volume} {361}},\ \bibinfo {pages}
  {387} (\bibinfo {year} {2018})}\BibitemShut {NoStop}%
\bibitem [{\citenamefont {Franklin}(2013)}]{Franklin2013}%
  \BibitemOpen
  \bibfield  {author} {\bibinfo {author} {\bibfnamefont {A.~D.}\ \bibnamefont
  {Franklin}},\ }\bibfield  {title} {\bibinfo {title} {The road to carbon
  nanotube transistors},\ }\href {\doibase 10.1038/498443a} {\bibfield
  {journal} {\bibinfo  {journal} {Nature}\ }\textbf {\bibinfo {volume} {498}},\
  \bibinfo {pages} {443} (\bibinfo {year} {2013})}\BibitemShut {NoStop}%
\bibitem [{\citenamefont {Brady}\ \emph {et~al.}(2016)\citenamefont {Brady},
  \citenamefont {Way}, \citenamefont {Safron}, \citenamefont {Evensen},
  \citenamefont {Gopalan},\ and\ \citenamefont {Arnold}}]{Brady2016}%
  \BibitemOpen
  \bibfield  {author} {\bibinfo {author} {\bibfnamefont {G.~J.}\ \bibnamefont
  {Brady}}, \bibinfo {author} {\bibfnamefont {A.~J.}\ \bibnamefont {Way}},
  \bibinfo {author} {\bibfnamefont {N.~S.}\ \bibnamefont {Safron}}, \bibinfo
  {author} {\bibfnamefont {H.~T.}\ \bibnamefont {Evensen}}, \bibinfo {author}
  {\bibfnamefont {P.}~\bibnamefont {Gopalan}}, \ and\ \bibinfo {author}
  {\bibfnamefont {M.~S.}\ \bibnamefont {Arnold}},\ }\bibfield  {title}
  {\bibinfo {title} {Quasi-ballistic carbon nanotube array transistors with
  current density exceeding {S}i and {GaAs}},\ }\href {\doibase
  10.1126/sciadv.1601240} {\bibfield  {journal} {\bibinfo  {journal} {Sci.
  Adv.}\ }\textbf {\bibinfo {volume} {2}},\ \bibinfo {pages} {e1601240}
  (\bibinfo {year} {2016})}\BibitemShut {NoStop}%
\bibitem [{\citenamefont {Liu}\ \emph {et~al.}(2020)\citenamefont {Liu},
  \citenamefont {Han}, \citenamefont {Xu}, \citenamefont {Zhou}, \citenamefont
  {Zhao}, \citenamefont {Ding}, \citenamefont {Shi}, \citenamefont {Xiao},
  \citenamefont {Ding}, \citenamefont {Ma}, \citenamefont {Jin}, \citenamefont
  {Zhang},\ and\ \citenamefont {Peng}}]{Liu2020}%
  \BibitemOpen
  \bibfield  {author} {\bibinfo {author} {\bibfnamefont {L.}~\bibnamefont
  {Liu}}, \bibinfo {author} {\bibfnamefont {J.}~\bibnamefont {Han}}, \bibinfo
  {author} {\bibfnamefont {L.}~\bibnamefont {Xu}}, \bibinfo {author}
  {\bibfnamefont {J.}~\bibnamefont {Zhou}}, \bibinfo {author} {\bibfnamefont
  {C.}~\bibnamefont {Zhao}}, \bibinfo {author} {\bibfnamefont {S.}~\bibnamefont
  {Ding}}, \bibinfo {author} {\bibfnamefont {H.}~\bibnamefont {Shi}}, \bibinfo
  {author} {\bibfnamefont {M.}~\bibnamefont {Xiao}}, \bibinfo {author}
  {\bibfnamefont {L.}~\bibnamefont {Ding}}, \bibinfo {author} {\bibfnamefont
  {Z.}~\bibnamefont {Ma}}, \bibinfo {author} {\bibfnamefont {C.}~\bibnamefont
  {Jin}}, \bibinfo {author} {\bibfnamefont {Z.}~\bibnamefont {Zhang}}, \ and\
  \bibinfo {author} {\bibfnamefont {L.-M.}\ \bibnamefont {Peng}},\ }\bibfield
  {title} {\bibinfo {title} {Aligned, high-density semiconducting carbon
  nanotube arrays for high-performance electronics},\ }\href {\doibase
  10.1126/science.aba5980} {\bibfield  {journal} {\bibinfo  {journal}
  {Science}\ }\textbf {\bibinfo {volume} {368}},\ \bibinfo {pages} {850}
  (\bibinfo {year} {2020})}\BibitemShut {NoStop}%
\bibitem [{\citenamefont {Jinkins}\ \emph {et~al.}(2021)\citenamefont
  {Jinkins}, \citenamefont {Foradori}, \citenamefont {Saraswat}, \citenamefont
  {Jacobberger}, \citenamefont {Dwyer}, \citenamefont {Gopalan}, \citenamefont
  {Berson},\ and\ \citenamefont {Arnold}}]{Jinkins2021}%
  \BibitemOpen
  \bibfield  {author} {\bibinfo {author} {\bibfnamefont {K.~R.}\ \bibnamefont
  {Jinkins}}, \bibinfo {author} {\bibfnamefont {S.~M.}\ \bibnamefont
  {Foradori}}, \bibinfo {author} {\bibfnamefont {V.}~\bibnamefont {Saraswat}},
  \bibinfo {author} {\bibfnamefont {R.~M.}\ \bibnamefont {Jacobberger}},
  \bibinfo {author} {\bibfnamefont {J.~H.}\ \bibnamefont {Dwyer}}, \bibinfo
  {author} {\bibfnamefont {P.}~\bibnamefont {Gopalan}}, \bibinfo {author}
  {\bibfnamefont {A.}~\bibnamefont {Berson}}, \ and\ \bibinfo {author}
  {\bibfnamefont {M.~S.}\ \bibnamefont {Arnold}},\ }\bibfield  {title}
  {\bibinfo {title} {Aligned {2D} carbon nanotube liquid crystals for
  wafer-scale electronics},\ }\href {\doibase 10.1126/sciadv.abh0640}
  {\bibfield  {journal} {\bibinfo  {journal} {Sci. Adv.}\ }\textbf {\bibinfo
  {volume} {7}},\ \bibinfo {pages} {eabh0640} (\bibinfo {year}
  {2021})}\BibitemShut {NoStop}%
\bibitem [{\citenamefont {Jin}\ \emph {et~al.}(2013)\citenamefont {Jin},
  \citenamefont {Dunham}, \citenamefont {Song}, \citenamefont {Xie},
  \citenamefont {hun Kim}, \citenamefont {Lu}, \citenamefont {Islam},
  \citenamefont {Du}, \citenamefont {Kim}, \citenamefont {Felts}, \citenamefont
  {Li}, \citenamefont {Xiong}, \citenamefont {Wahab}, \citenamefont {Menon},
  \citenamefont {Cho}, \citenamefont {Grosse}, \citenamefont {Lee},
  \citenamefont {Chung}, \citenamefont {Pop}, \citenamefont {Alam},
  \citenamefont {King}, \citenamefont {Huang},\ and\ \citenamefont
  {Rogers}}]{Jin2013}%
  \BibitemOpen
  \bibfield  {author} {\bibinfo {author} {\bibfnamefont {S.~H.}\ \bibnamefont
  {Jin}}, \bibinfo {author} {\bibfnamefont {S.~N.}\ \bibnamefont {Dunham}},
  \bibinfo {author} {\bibfnamefont {J.}~\bibnamefont {Song}}, \bibinfo {author}
  {\bibfnamefont {X.}~\bibnamefont {Xie}}, \bibinfo {author} {\bibfnamefont
  {J.}~\bibnamefont {hun Kim}}, \bibinfo {author} {\bibfnamefont
  {C.}~\bibnamefont {Lu}}, \bibinfo {author} {\bibfnamefont {A.}~\bibnamefont
  {Islam}}, \bibinfo {author} {\bibfnamefont {F.}~\bibnamefont {Du}}, \bibinfo
  {author} {\bibfnamefont {J.}~\bibnamefont {Kim}}, \bibinfo {author}
  {\bibfnamefont {J.}~\bibnamefont {Felts}}, \bibinfo {author} {\bibfnamefont
  {Y.}~\bibnamefont {Li}}, \bibinfo {author} {\bibfnamefont {F.}~\bibnamefont
  {Xiong}}, \bibinfo {author} {\bibfnamefont {M.~A.}\ \bibnamefont {Wahab}},
  \bibinfo {author} {\bibfnamefont {M.}~\bibnamefont {Menon}}, \bibinfo
  {author} {\bibfnamefont {E.}~\bibnamefont {Cho}}, \bibinfo {author}
  {\bibfnamefont {K.~L.}\ \bibnamefont {Grosse}}, \bibinfo {author}
  {\bibfnamefont {D.~J.}\ \bibnamefont {Lee}}, \bibinfo {author} {\bibfnamefont
  {H.~U.}\ \bibnamefont {Chung}}, \bibinfo {author} {\bibfnamefont
  {E.}~\bibnamefont {Pop}}, \bibinfo {author} {\bibfnamefont {M.~A.}\
  \bibnamefont {Alam}}, \bibinfo {author} {\bibfnamefont {W.~P.}\ \bibnamefont
  {King}}, \bibinfo {author} {\bibfnamefont {Y.}~\bibnamefont {Huang}}, \ and\
  \bibinfo {author} {\bibfnamefont {J.~A.}\ \bibnamefont {Rogers}},\ }\bibfield
   {title} {\bibinfo {title} {Using nanoscale thermocapillary flows to create
  arrays of purely semiconducting single-walled carbon nanotubes},\ }\href
  {\doibase 10.1038/nnano.2013.56} {\bibfield  {journal} {\bibinfo  {journal}
  {Nat. Nanotechnol.}\ }\textbf {\bibinfo {volume} {8}},\ \bibinfo {pages}
  {347} (\bibinfo {year} {2013})}\BibitemShut {NoStop}%
\bibitem [{\citenamefont {Shulaker}\ \emph {et~al.}(2017)\citenamefont
  {Shulaker}, \citenamefont {Hills}, \citenamefont {Park}, \citenamefont
  {Howe}, \citenamefont {Saraswat}, \citenamefont {Wong},\ and\ \citenamefont
  {Mitra}}]{Shulaker2017}%
  \BibitemOpen
  \bibfield  {author} {\bibinfo {author} {\bibfnamefont {M.~M.}\ \bibnamefont
  {Shulaker}}, \bibinfo {author} {\bibfnamefont {G.}~\bibnamefont {Hills}},
  \bibinfo {author} {\bibfnamefont {R.~S.}\ \bibnamefont {Park}}, \bibinfo
  {author} {\bibfnamefont {R.~T.}\ \bibnamefont {Howe}}, \bibinfo {author}
  {\bibfnamefont {K.}~\bibnamefont {Saraswat}}, \bibinfo {author}
  {\bibfnamefont {H.-S.~P.}\ \bibnamefont {Wong}}, \ and\ \bibinfo {author}
  {\bibfnamefont {S.}~\bibnamefont {Mitra}},\ }\bibfield  {title} {\bibinfo
  {title} {Three-dimensional integration of nanotechnologies for computing and
  data storage on a single chip},\ }\href {\doibase 10.1038/nature22994}
  {\bibfield  {journal} {\bibinfo  {journal} {Nature}\ }\textbf {\bibinfo
  {volume} {547}},\ \bibinfo {pages} {74} (\bibinfo {year} {2017})}\BibitemShut
  {NoStop}%
\bibitem [{\citenamefont {Otsuka}\ \emph {et~al.}(2017)\citenamefont {Otsuka},
  \citenamefont {Inoue}, \citenamefont {Maeda}, \citenamefont {Kometani},
  \citenamefont {Chiashi},\ and\ \citenamefont {Maruyama}}]{Otsuka2017}%
  \BibitemOpen
  \bibfield  {author} {\bibinfo {author} {\bibfnamefont {K.}~\bibnamefont
  {Otsuka}}, \bibinfo {author} {\bibfnamefont {T.}~\bibnamefont {Inoue}},
  \bibinfo {author} {\bibfnamefont {E.}~\bibnamefont {Maeda}}, \bibinfo
  {author} {\bibfnamefont {R.}~\bibnamefont {Kometani}}, \bibinfo {author}
  {\bibfnamefont {S.}~\bibnamefont {Chiashi}}, \ and\ \bibinfo {author}
  {\bibfnamefont {S.}~\bibnamefont {Maruyama}},\ }\bibfield  {title} {\bibinfo
  {title} {On-chip sorting of long semiconducting carbon nanotubes for multiple
  transistors along an identical array},\ }\href {\doibase
  10.1021/acsnano.7b06282} {\bibfield  {journal} {\bibinfo  {journal} {ACS
  Nano}\ }\textbf {\bibinfo {volume} {11}},\ \bibinfo {pages} {11497} (\bibinfo
  {year} {2017})}\BibitemShut {NoStop}%
\bibitem [{\citenamefont {Hong}\ \emph {et~al.}(2010)\citenamefont {Hong},
  \citenamefont {Banks},\ and\ \citenamefont {Rogers}}]{Hong2010}%
  \BibitemOpen
  \bibfield  {author} {\bibinfo {author} {\bibfnamefont {S.~W.}\ \bibnamefont
  {Hong}}, \bibinfo {author} {\bibfnamefont {T.}~\bibnamefont {Banks}}, \ and\
  \bibinfo {author} {\bibfnamefont {J.~A.}\ \bibnamefont {Rogers}},\ }\bibfield
   {title} {\bibinfo {title} {Improved density in aligned arrays of
  single-walled carbon nanotubes by sequential chemical vapor deposition on
  quartz},\ }\href {\doibase 10.1002/adma.200903238} {\bibfield  {journal}
  {\bibinfo  {journal} {Adv. Mater.}\ }\textbf {\bibinfo {volume} {22}},\
  \bibinfo {pages} {1826} (\bibinfo {year} {2010})}\BibitemShut {NoStop}%
\bibitem [{\citenamefont {Hu}\ \emph {et~al.}(2015)\citenamefont {Hu},
  \citenamefont {Kang}, \citenamefont {Zhao}, \citenamefont {Zhong},
  \citenamefont {Zhang}, \citenamefont {Yang}, \citenamefont {Wang},
  \citenamefont {Lin}, \citenamefont {Li}, \citenamefont {Zhang}, \citenamefont
  {Peng}, \citenamefont {Liu},\ and\ \citenamefont {Zhang}}]{Hu2015}%
  \BibitemOpen
  \bibfield  {author} {\bibinfo {author} {\bibfnamefont {Y.}~\bibnamefont
  {Hu}}, \bibinfo {author} {\bibfnamefont {L.}~\bibnamefont {Kang}}, \bibinfo
  {author} {\bibfnamefont {Q.}~\bibnamefont {Zhao}}, \bibinfo {author}
  {\bibfnamefont {H.}~\bibnamefont {Zhong}}, \bibinfo {author} {\bibfnamefont
  {S.}~\bibnamefont {Zhang}}, \bibinfo {author} {\bibfnamefont
  {L.}~\bibnamefont {Yang}}, \bibinfo {author} {\bibfnamefont {Z.}~\bibnamefont
  {Wang}}, \bibinfo {author} {\bibfnamefont {J.}~\bibnamefont {Lin}}, \bibinfo
  {author} {\bibfnamefont {Q.}~\bibnamefont {Li}}, \bibinfo {author}
  {\bibfnamefont {Z.}~\bibnamefont {Zhang}}, \bibinfo {author} {\bibfnamefont
  {L.}~\bibnamefont {Peng}}, \bibinfo {author} {\bibfnamefont {Z.}~\bibnamefont
  {Liu}}, \ and\ \bibinfo {author} {\bibfnamefont {J.}~\bibnamefont {Zhang}},\
  }\bibfield  {title} {\bibinfo {title} {Growth of high-density horizontally
  aligned {SWNT} arrays using trojan catalysts},\ }\href {\doibase
  10.1038/ncomms7099} {\bibfield  {journal} {\bibinfo  {journal} {Nat.
  Commun.}\ }\textbf {\bibinfo {volume} {6}},\ \bibinfo {pages} {6099}
  (\bibinfo {year} {2015})}\BibitemShut {NoStop}%
\bibitem [{\citenamefont {Xu}\ \emph {et~al.}(2018)\citenamefont {Xu},
  \citenamefont {Gao}, \citenamefont {Zhang},\ and\ \citenamefont
  {Peng}}]{Xu2018}%
  \BibitemOpen
  \bibfield  {author} {\bibinfo {author} {\bibfnamefont {L.}~\bibnamefont
  {Xu}}, \bibinfo {author} {\bibfnamefont {N.}~\bibnamefont {Gao}}, \bibinfo
  {author} {\bibfnamefont {Z.}~\bibnamefont {Zhang}}, \ and\ \bibinfo {author}
  {\bibfnamefont {L.-M.}\ \bibnamefont {Peng}},\ }\bibfield  {title} {\bibinfo
  {title} {Lowering interface state density in carbon nanotube thin film
  transistors through using stacked {Y}$_2${O}$_3$/{HfO}$_2$ gate dielectric},\
  }\href {\doibase 10.1063/1.5039967} {\bibfield  {journal} {\bibinfo
  {journal} {Appl. Phys. Lett.}\ }\textbf {\bibinfo {volume} {113}},\ \bibinfo
  {pages} {083105} (\bibinfo {year} {2018})}\BibitemShut {NoStop}%
\bibitem [{\citenamefont {Ding}\ \emph
  {et~al.}(2009{\natexlab{a}})\citenamefont {Ding}, \citenamefont
  {Harutyunyan},\ and\ \citenamefont {Yakobson}}]{Ding2009}%
  \BibitemOpen
  \bibfield  {author} {\bibinfo {author} {\bibfnamefont {F.}~\bibnamefont
  {Ding}}, \bibinfo {author} {\bibfnamefont {A.~R.}\ \bibnamefont
  {Harutyunyan}}, \ and\ \bibinfo {author} {\bibfnamefont {B.~I.}\ \bibnamefont
  {Yakobson}},\ }\bibfield  {title} {\bibinfo {title} {Dislocation theory of
  chirality-controlled nanotube growth},\ }\href {\doibase
  10.1073/pnas.0811946106} {\bibfield  {journal} {\bibinfo  {journal} {Proc.
  Natl. Acad. Sci. U.S.A.}\ }\textbf {\bibinfo {volume} {106}},\ \bibinfo
  {pages} {2506} (\bibinfo {year} {2009}{\natexlab{a}})}\BibitemShut {NoStop}%
\bibitem [{\citenamefont {Artyukhov}\ \emph {et~al.}(2014)\citenamefont
  {Artyukhov}, \citenamefont {Penev},\ and\ \citenamefont
  {Yakobson}}]{Artyukhov2014}%
  \BibitemOpen
  \bibfield  {author} {\bibinfo {author} {\bibfnamefont {V.~I.}\ \bibnamefont
  {Artyukhov}}, \bibinfo {author} {\bibfnamefont {E.~S.}\ \bibnamefont
  {Penev}}, \ and\ \bibinfo {author} {\bibfnamefont {B.~I.}\ \bibnamefont
  {Yakobson}},\ }\bibfield  {title} {\bibinfo {title} {Why nanotubes grow
  chiral},\ }\href {\doibase 10.1038/ncomms5892} {\bibfield  {journal}
  {\bibinfo  {journal} {Nat. Commun.}\ }\textbf {\bibinfo {volume} {5}},\
  \bibinfo {pages} {4892} (\bibinfo {year} {2014})}\BibitemShut {NoStop}%
\bibitem [{\citenamefont {Yang}\ \emph {et~al.}(2014)\citenamefont {Yang},
  \citenamefont {Wang}, \citenamefont {Zhang}, \citenamefont {Yang},
  \citenamefont {Luo}, \citenamefont {Xu}, \citenamefont {Wei}, \citenamefont
  {Wang}, \citenamefont {Xu}, \citenamefont {Peng}, \citenamefont {Li},
  \citenamefont {Li}, \citenamefont {Li}, \citenamefont {Li}, \citenamefont
  {Bai}, \citenamefont {Ding},\ and\ \citenamefont {Li}}]{Yang2014}%
  \BibitemOpen
  \bibfield  {author} {\bibinfo {author} {\bibfnamefont {F.}~\bibnamefont
  {Yang}}, \bibinfo {author} {\bibfnamefont {X.}~\bibnamefont {Wang}}, \bibinfo
  {author} {\bibfnamefont {D.}~\bibnamefont {Zhang}}, \bibinfo {author}
  {\bibfnamefont {J.}~\bibnamefont {Yang}}, \bibinfo {author} {\bibfnamefont
  {D.}~\bibnamefont {Luo}}, \bibinfo {author} {\bibfnamefont {Z.}~\bibnamefont
  {Xu}}, \bibinfo {author} {\bibfnamefont {J.}~\bibnamefont {Wei}}, \bibinfo
  {author} {\bibfnamefont {J.-Q.}\ \bibnamefont {Wang}}, \bibinfo {author}
  {\bibfnamefont {Z.}~\bibnamefont {Xu}}, \bibinfo {author} {\bibfnamefont
  {F.}~\bibnamefont {Peng}}, \bibinfo {author} {\bibfnamefont {X.}~\bibnamefont
  {Li}}, \bibinfo {author} {\bibfnamefont {R.}~\bibnamefont {Li}}, \bibinfo
  {author} {\bibfnamefont {Y.}~\bibnamefont {Li}}, \bibinfo {author}
  {\bibfnamefont {M.}~\bibnamefont {Li}}, \bibinfo {author} {\bibfnamefont
  {X.}~\bibnamefont {Bai}}, \bibinfo {author} {\bibfnamefont {F.}~\bibnamefont
  {Ding}}, \ and\ \bibinfo {author} {\bibfnamefont {Y.}~\bibnamefont {Li}},\
  }\bibfield  {title} {\bibinfo {title} {Chirality-specific growth of
  single-walled carbon nanotubes on solid alloy catalysts},\ }\href {\doibase
  10.1038/nature13434} {\bibfield  {journal} {\bibinfo  {journal} {Nature}\
  }\textbf {\bibinfo {volume} {510}},\ \bibinfo {pages} {522} (\bibinfo {year}
  {2014})}\BibitemShut {NoStop}%
\bibitem [{\citenamefont {Zhang}\ \emph {et~al.}(2017)\citenamefont {Zhang},
  \citenamefont {Kang}, \citenamefont {Wang}, \citenamefont {Tong},
  \citenamefont {Yang}, \citenamefont {Wang}, \citenamefont {Qi}, \citenamefont
  {Deng}, \citenamefont {Li}, \citenamefont {Bai}, \citenamefont {Ding},\ and\
  \citenamefont {Zhang}}]{Zhang2017}%
  \BibitemOpen
  \bibfield  {author} {\bibinfo {author} {\bibfnamefont {S.}~\bibnamefont
  {Zhang}}, \bibinfo {author} {\bibfnamefont {L.}~\bibnamefont {Kang}},
  \bibinfo {author} {\bibfnamefont {X.}~\bibnamefont {Wang}}, \bibinfo {author}
  {\bibfnamefont {L.}~\bibnamefont {Tong}}, \bibinfo {author} {\bibfnamefont
  {L.}~\bibnamefont {Yang}}, \bibinfo {author} {\bibfnamefont {Z.}~\bibnamefont
  {Wang}}, \bibinfo {author} {\bibfnamefont {K.}~\bibnamefont {Qi}}, \bibinfo
  {author} {\bibfnamefont {S.}~\bibnamefont {Deng}}, \bibinfo {author}
  {\bibfnamefont {Q.}~\bibnamefont {Li}}, \bibinfo {author} {\bibfnamefont
  {X.}~\bibnamefont {Bai}}, \bibinfo {author} {\bibfnamefont {F.}~\bibnamefont
  {Ding}}, \ and\ \bibinfo {author} {\bibfnamefont {J.}~\bibnamefont {Zhang}},\
  }\bibfield  {title} {\bibinfo {title} {Arrays of horizontal carbon nanotubes
  of controlled chirality grown using designed catalysts},\ }\href {\doibase
  10.1038/nature21051} {\bibfield  {journal} {\bibinfo  {journal} {Nature}\
  }\textbf {\bibinfo {volume} {543}},\ \bibinfo {pages} {234} (\bibinfo {year}
  {2017})}\BibitemShut {NoStop}%
\bibitem [{\citenamefont {Zhu}\ \emph {et~al.}(2019)\citenamefont {Zhu},
  \citenamefont {Wei}, \citenamefont {Cheng}, \citenamefont {Shen},
  \citenamefont {Sun}, \citenamefont {Gao}, \citenamefont {Wen}, \citenamefont
  {Zhang}, \citenamefont {Xu}, \citenamefont {Wang},\ and\ \citenamefont
  {Wei}}]{Zhu2019}%
  \BibitemOpen
  \bibfield  {author} {\bibinfo {author} {\bibfnamefont {Z.}~\bibnamefont
  {Zhu}}, \bibinfo {author} {\bibfnamefont {N.}~\bibnamefont {Wei}}, \bibinfo
  {author} {\bibfnamefont {W.}~\bibnamefont {Cheng}}, \bibinfo {author}
  {\bibfnamefont {B.}~\bibnamefont {Shen}}, \bibinfo {author} {\bibfnamefont
  {S.}~\bibnamefont {Sun}}, \bibinfo {author} {\bibfnamefont {J.}~\bibnamefont
  {Gao}}, \bibinfo {author} {\bibfnamefont {Q.}~\bibnamefont {Wen}}, \bibinfo
  {author} {\bibfnamefont {R.}~\bibnamefont {Zhang}}, \bibinfo {author}
  {\bibfnamefont {J.}~\bibnamefont {Xu}}, \bibinfo {author} {\bibfnamefont
  {Y.}~\bibnamefont {Wang}}, \ and\ \bibinfo {author} {\bibfnamefont
  {F.}~\bibnamefont {Wei}},\ }\bibfield  {title} {\bibinfo {title}
  {Rate-selected growth of ultrapure semiconducting carbon nanotube arrays},\
  }\href {\doibase 10.1038/s41467-019-12519-5} {\bibfield  {journal} {\bibinfo
  {journal} {Nat. Commun.}\ }\textbf {\bibinfo {volume} {10}},\ \bibinfo
  {pages} {4467} (\bibinfo {year} {2019})}\BibitemShut {NoStop}%
\bibitem [{\citenamefont {Yang}\ \emph {et~al.}(2020)\citenamefont {Yang},
  \citenamefont {Wang}, \citenamefont {Zhang}, \citenamefont {Yang},
  \citenamefont {Zheng},\ and\ \citenamefont {Li}}]{Yang2020}%
  \BibitemOpen
  \bibfield  {author} {\bibinfo {author} {\bibfnamefont {F.}~\bibnamefont
  {Yang}}, \bibinfo {author} {\bibfnamefont {M.}~\bibnamefont {Wang}}, \bibinfo
  {author} {\bibfnamefont {D.}~\bibnamefont {Zhang}}, \bibinfo {author}
  {\bibfnamefont {J.}~\bibnamefont {Yang}}, \bibinfo {author} {\bibfnamefont
  {M.}~\bibnamefont {Zheng}}, \ and\ \bibinfo {author} {\bibfnamefont
  {Y.}~\bibnamefont {Li}},\ }\bibfield  {title} {\bibinfo {title} {Chirality
  pure carbon nanotubes: Growth, sorting, and characterization},\ }\href
  {\doibase 10.1021/acs.chemrev.9b00835} {\bibfield  {journal} {\bibinfo
  {journal} {Chem. Rev.}\ }\textbf {\bibinfo {volume} {120}},\ \bibinfo {pages}
  {2693} (\bibinfo {year} {2020})}\BibitemShut {NoStop}%
\bibitem [{\citenamefont {Förster}\ \emph {et~al.}(2021)\citenamefont
  {Förster}, \citenamefont {Swinburne}, \citenamefont {Jiang}, \citenamefont
  {Kauppinen},\ and\ \citenamefont {Bichara}}]{Foerster2021}%
  \BibitemOpen
  \bibfield  {author} {\bibinfo {author} {\bibfnamefont {G.~D.}\ \bibnamefont
  {Förster}}, \bibinfo {author} {\bibfnamefont {T.~D.}\ \bibnamefont
  {Swinburne}}, \bibinfo {author} {\bibfnamefont {H.}~\bibnamefont {Jiang}},
  \bibinfo {author} {\bibfnamefont {E.}~\bibnamefont {Kauppinen}}, \ and\
  \bibinfo {author} {\bibfnamefont {C.}~\bibnamefont {Bichara}},\ }\bibfield
  {title} {\bibinfo {title} {A semi-grand canonical kinetic {M}onte {C}arlo
  study of single-walled carbon nanotube growth},\ }\href {\doibase
  10.1063/5.0030943} {\bibfield  {journal} {\bibinfo  {journal} {AIP Adv.}\
  }\textbf {\bibinfo {volume} {11}},\ \bibinfo {pages} {045306} (\bibinfo
  {year} {2021})}\BibitemShut {NoStop}%
\bibitem [{\citenamefont {Inoue}\ \emph {et~al.}(2015)\citenamefont {Inoue},
  \citenamefont {Hasegawa}, \citenamefont {Chiashi},\ and\ \citenamefont
  {Maruyama}}]{Inoue2015}%
  \BibitemOpen
  \bibfield  {author} {\bibinfo {author} {\bibfnamefont {T.}~\bibnamefont
  {Inoue}}, \bibinfo {author} {\bibfnamefont {D.}~\bibnamefont {Hasegawa}},
  \bibinfo {author} {\bibfnamefont {S.}~\bibnamefont {Chiashi}}, \ and\
  \bibinfo {author} {\bibfnamefont {S.}~\bibnamefont {Maruyama}},\ }\bibfield
  {title} {\bibinfo {title} {Chirality analysis of horizontally aligned
  single-walled carbon nanotubes: decoupling populations and lengths},\ }\href
  {\doibase 10.1039/c5ta02679b} {\bibfield  {journal} {\bibinfo  {journal} {J.
  Mater. Chem.A}\ }\textbf {\bibinfo {volume} {3}},\ \bibinfo {pages} {15119}
  (\bibinfo {year} {2015})}\BibitemShut {NoStop}%
\bibitem [{\citenamefont {Rao}\ \emph {et~al.}(2012)\citenamefont {Rao},
  \citenamefont {Liptak}, \citenamefont {Cherukuri}, \citenamefont {Yakobson},\
  and\ \citenamefont {Maruyama}}]{Rao2012}%
  \BibitemOpen
  \bibfield  {author} {\bibinfo {author} {\bibfnamefont {R.}~\bibnamefont
  {Rao}}, \bibinfo {author} {\bibfnamefont {D.}~\bibnamefont {Liptak}},
  \bibinfo {author} {\bibfnamefont {T.}~\bibnamefont {Cherukuri}}, \bibinfo
  {author} {\bibfnamefont {B.~I.}\ \bibnamefont {Yakobson}}, \ and\ \bibinfo
  {author} {\bibfnamefont {B.}~\bibnamefont {Maruyama}},\ }\bibfield  {title}
  {\bibinfo {title} {In situ evidence for chirality-dependent growth rates of
  individual carbon nanotubes},\ }\href {\doibase 10.1038/nmat3231} {\bibfield
  {journal} {\bibinfo  {journal} {Nat. Mater.}\ }\textbf {\bibinfo {volume}
  {11}},\ \bibinfo {pages} {213} (\bibinfo {year} {2012})}\BibitemShut
  {NoStop}%
\bibitem [{\citenamefont {Otsuka}\ \emph {et~al.}(2018)\citenamefont {Otsuka},
  \citenamefont {Yamamoto}, \citenamefont {Inoue}, \citenamefont {Koyano},
  \citenamefont {Ukai}, \citenamefont {Yoshikawa}, \citenamefont {Xiang},
  \citenamefont {Chiashi},\ and\ \citenamefont {Maruyama}}]{Otsuka2018}%
  \BibitemOpen
  \bibfield  {author} {\bibinfo {author} {\bibfnamefont {K.}~\bibnamefont
  {Otsuka}}, \bibinfo {author} {\bibfnamefont {S.}~\bibnamefont {Yamamoto}},
  \bibinfo {author} {\bibfnamefont {T.}~\bibnamefont {Inoue}}, \bibinfo
  {author} {\bibfnamefont {B.}~\bibnamefont {Koyano}}, \bibinfo {author}
  {\bibfnamefont {H.}~\bibnamefont {Ukai}}, \bibinfo {author} {\bibfnamefont
  {R.}~\bibnamefont {Yoshikawa}}, \bibinfo {author} {\bibfnamefont
  {R.}~\bibnamefont {Xiang}}, \bibinfo {author} {\bibfnamefont
  {S.}~\bibnamefont {Chiashi}}, \ and\ \bibinfo {author} {\bibfnamefont
  {S.}~\bibnamefont {Maruyama}},\ }\bibfield  {title} {\bibinfo {title}
  {Digital isotope coding to trace the growth process of individual
  single-walled carbon nanotubes},\ }\href {\doibase 10.1021/acsnano.8b01630}
  {\bibfield  {journal} {\bibinfo  {journal} {ACS Nano}\ }\textbf {\bibinfo
  {volume} {12}},\ \bibinfo {pages} {3994} (\bibinfo {year}
  {2018})}\BibitemShut {NoStop}%
\bibitem [{\citenamefont {He}\ \emph {et~al.}(2019)\citenamefont {He},
  \citenamefont {Wang}, \citenamefont {Zhang}, \citenamefont {Jiang},
  \citenamefont {Cavalca}, \citenamefont {Cui}, \citenamefont {Wagner},
  \citenamefont {Hansen}, \citenamefont {Kauppinen}, \citenamefont {Zhang},\
  and\ \citenamefont {Ding}}]{He2019}%
  \BibitemOpen
  \bibfield  {author} {\bibinfo {author} {\bibfnamefont {M.}~\bibnamefont
  {He}}, \bibinfo {author} {\bibfnamefont {X.}~\bibnamefont {Wang}}, \bibinfo
  {author} {\bibfnamefont {S.}~\bibnamefont {Zhang}}, \bibinfo {author}
  {\bibfnamefont {H.}~\bibnamefont {Jiang}}, \bibinfo {author} {\bibfnamefont
  {F.}~\bibnamefont {Cavalca}}, \bibinfo {author} {\bibfnamefont
  {H.}~\bibnamefont {Cui}}, \bibinfo {author} {\bibfnamefont {J.~B.}\
  \bibnamefont {Wagner}}, \bibinfo {author} {\bibfnamefont {T.~W.}\
  \bibnamefont {Hansen}}, \bibinfo {author} {\bibfnamefont {E.}~\bibnamefont
  {Kauppinen}}, \bibinfo {author} {\bibfnamefont {J.}~\bibnamefont {Zhang}}, \
  and\ \bibinfo {author} {\bibfnamefont {F.}~\bibnamefont {Ding}},\ }\bibfield
  {title} {\bibinfo {title} {Growth kinetics of single-walled carbon nanotubes
  with a (2n,n) chirality selection},\ }\href {\doibase 10.1126/sciadv.aav9668}
  {\bibfield  {journal} {\bibinfo  {journal} {Sci. Adv.}\ }\textbf {\bibinfo
  {volume} {5}},\ \bibinfo {pages} {eaav9668} (\bibinfo {year}
  {2019})}\BibitemShut {NoStop}%
\bibitem [{\citenamefont {Pimonov}\ \emph {et~al.}(2021)\citenamefont
  {Pimonov}, \citenamefont {Tran}, \citenamefont {Monniello}, \citenamefont
  {Tahir}, \citenamefont {Michel}, \citenamefont {Podor}, \citenamefont
  {Odorico}, \citenamefont {Bichara},\ and\ \citenamefont
  {Jourdain}}]{Pimonov2021}%
  \BibitemOpen
  \bibfield  {author} {\bibinfo {author} {\bibfnamefont {V.}~\bibnamefont
  {Pimonov}}, \bibinfo {author} {\bibfnamefont {H.-N.}\ \bibnamefont {Tran}},
  \bibinfo {author} {\bibfnamefont {L.}~\bibnamefont {Monniello}}, \bibinfo
  {author} {\bibfnamefont {S.}~\bibnamefont {Tahir}}, \bibinfo {author}
  {\bibfnamefont {T.}~\bibnamefont {Michel}}, \bibinfo {author} {\bibfnamefont
  {R.}~\bibnamefont {Podor}}, \bibinfo {author} {\bibfnamefont
  {M.}~\bibnamefont {Odorico}}, \bibinfo {author} {\bibfnamefont
  {C.}~\bibnamefont {Bichara}}, \ and\ \bibinfo {author} {\bibfnamefont
  {V.}~\bibnamefont {Jourdain}},\ }\bibfield  {title} {\bibinfo {title}
  {Dynamic instability of individual carbon nanotube growth revealed by in situ
  homodyne polarization microscopy},\ }\href {\doibase
  10.1021/acs.nanolett.1c03431} {\bibfield  {journal} {\bibinfo  {journal}
  {Nano Lett.}\ }\textbf {\bibinfo {volume} {21}},\ \bibinfo {pages} {8495}
  (\bibinfo {year} {2021})}\BibitemShut {NoStop}%
\bibitem [{\citenamefont {Maruyama}\ \emph {et~al.}(2002)\citenamefont
  {Maruyama}, \citenamefont {Kojima}, \citenamefont {Miyauchi}, \citenamefont
  {Chiashi},\ and\ \citenamefont {Kohno}}]{Maruyama2002}%
  \BibitemOpen
  \bibfield  {author} {\bibinfo {author} {\bibfnamefont {S.}~\bibnamefont
  {Maruyama}}, \bibinfo {author} {\bibfnamefont {R.}~\bibnamefont {Kojima}},
  \bibinfo {author} {\bibfnamefont {Y.}~\bibnamefont {Miyauchi}}, \bibinfo
  {author} {\bibfnamefont {S.}~\bibnamefont {Chiashi}}, \ and\ \bibinfo
  {author} {\bibfnamefont {M.}~\bibnamefont {Kohno}},\ }\bibfield  {title}
  {\bibinfo {title} {Low-temperature synthesis of high-purity single-walled
  carbon nanotubes from alcohol},\ }\href {\doibase
  10.1016/s0009-2614(02)00838-2} {\bibfield  {journal} {\bibinfo  {journal}
  {Chem. Phys. Lett.}\ }\textbf {\bibinfo {volume} {360}},\ \bibinfo {pages}
  {229} (\bibinfo {year} {2002})}\BibitemShut {NoStop}%
\bibitem [{\citenamefont {Zhou}\ \emph {et~al.}(2012)\citenamefont {Zhou},
  \citenamefont {Zhan}, \citenamefont {Ding},\ and\ \citenamefont
  {Liu}}]{Zhou2012}%
  \BibitemOpen
  \bibfield  {author} {\bibinfo {author} {\bibfnamefont {W.}~\bibnamefont
  {Zhou}}, \bibinfo {author} {\bibfnamefont {S.}~\bibnamefont {Zhan}}, \bibinfo
  {author} {\bibfnamefont {L.}~\bibnamefont {Ding}}, \ and\ \bibinfo {author}
  {\bibfnamefont {J.}~\bibnamefont {Liu}},\ }\bibfield  {title} {\bibinfo
  {title} {General rules for selective growth of enriched semiconducting single
  walled carbon nanotubes with water vapor as in situ etchant},\ }\href
  {\doibase 10.1021/ja3038992} {\bibfield  {journal} {\bibinfo  {journal} {J.
  Am. Chem. Soc.}\ }\textbf {\bibinfo {volume} {134}},\ \bibinfo {pages}
  {14019} (\bibinfo {year} {2012})}\BibitemShut {NoStop}%
\bibitem [{\citenamefont {Xiang}\ \emph {et~al.}(2010)\citenamefont {Xiang},
  \citenamefont {Einarsson}, \citenamefont {Okawa}, \citenamefont
  {Thurakitseree}, \citenamefont {Murakami}, \citenamefont {Shiomi},
  \citenamefont {Ohno},\ and\ \citenamefont {Maruyama}}]{Xiang2010}%
  \BibitemOpen
  \bibfield  {author} {\bibinfo {author} {\bibfnamefont {R.}~\bibnamefont
  {Xiang}}, \bibinfo {author} {\bibfnamefont {E.}~\bibnamefont {Einarsson}},
  \bibinfo {author} {\bibfnamefont {J.}~\bibnamefont {Okawa}}, \bibinfo
  {author} {\bibfnamefont {T.}~\bibnamefont {Thurakitseree}}, \bibinfo {author}
  {\bibfnamefont {Y.}~\bibnamefont {Murakami}}, \bibinfo {author}
  {\bibfnamefont {J.}~\bibnamefont {Shiomi}}, \bibinfo {author} {\bibfnamefont
  {Y.}~\bibnamefont {Ohno}}, \ and\ \bibinfo {author} {\bibfnamefont
  {S.}~\bibnamefont {Maruyama}},\ }\bibfield  {title} {\bibinfo {title}
  {Parametric study of alcohol catalytic chemical vapor deposition for
  controlled synthesis of vertically aligned single-walled carbon nanotubes},\
  }\href {\doibase 10.1166/jnn.2010.2011} {\bibfield  {journal} {\bibinfo
  {journal} {J. Nanosci. Nanotechnol.}\ }\textbf {\bibinfo {volume} {10}},\
  \bibinfo {pages} {3901} (\bibinfo {year} {2010})}\BibitemShut {NoStop}%
\bibitem [{\citenamefont {Minakov}\ \emph {et~al.}(2019)\citenamefont
  {Minakov}, \citenamefont {Simunin},\ and\ \citenamefont
  {Ryzhkov}}]{Minakov2019}%
  \BibitemOpen
  \bibfield  {author} {\bibinfo {author} {\bibfnamefont {A.~V.}\ \bibnamefont
  {Minakov}}, \bibinfo {author} {\bibfnamefont {M.~M.}\ \bibnamefont
  {Simunin}}, \ and\ \bibinfo {author} {\bibfnamefont {I.~I.}\ \bibnamefont
  {Ryzhkov}},\ }\bibfield  {title} {\bibinfo {title} {Modelling of ethanol
  pyrolysis in a commercial {CVD} reactor for growing carbon layers on alumina
  substrates},\ }\href {\doibase 10.1016/j.ijheatmasstransfer.2019.118764}
  {\bibfield  {journal} {\bibinfo  {journal} {Int. J. Heat Mass Transfer}\
  }\textbf {\bibinfo {volume} {145}},\ \bibinfo {pages} {118764} (\bibinfo
  {year} {2019})}\BibitemShut {NoStop}%
\bibitem [{\citenamefont {Feng}\ \emph {et~al.}(2011)\citenamefont {Feng},
  \citenamefont {Chee}, \citenamefont {Sharma}, \citenamefont {Liu},
  \citenamefont {Xie}, \citenamefont {Li}, \citenamefont {Fan},\ and\
  \citenamefont {Jiang}}]{Feng2011}%
  \BibitemOpen
  \bibfield  {author} {\bibinfo {author} {\bibfnamefont {X.}~\bibnamefont
  {Feng}}, \bibinfo {author} {\bibfnamefont {S.~W.}\ \bibnamefont {Chee}},
  \bibinfo {author} {\bibfnamefont {R.}~\bibnamefont {Sharma}}, \bibinfo
  {author} {\bibfnamefont {K.}~\bibnamefont {Liu}}, \bibinfo {author}
  {\bibfnamefont {X.}~\bibnamefont {Xie}}, \bibinfo {author} {\bibfnamefont
  {Q.}~\bibnamefont {Li}}, \bibinfo {author} {\bibfnamefont {S.}~\bibnamefont
  {Fan}}, \ and\ \bibinfo {author} {\bibfnamefont {K.}~\bibnamefont {Jiang}},\
  }\bibfield  {title} {\bibinfo {title} {In situ {TEM} observation of the
  gasification and growth of carbon nanotubes using iron catalysts},\ }\href
  {\doibase 10.1007/s12274-011-0133-x} {\bibfield  {journal} {\bibinfo
  {journal} {Nano Res.}\ }\textbf {\bibinfo {volume} {4}},\ \bibinfo {pages}
  {767} (\bibinfo {year} {2011})}\BibitemShut {NoStop}%
\bibitem [{\citenamefont {Ishii}\ \emph {et~al.}(2015)\citenamefont {Ishii},
  \citenamefont {Yoshida},\ and\ \citenamefont {Kato}}]{Ishii2015}%
  \BibitemOpen
  \bibfield  {author} {\bibinfo {author} {\bibfnamefont {A.}~\bibnamefont
  {Ishii}}, \bibinfo {author} {\bibfnamefont {M.}~\bibnamefont {Yoshida}}, \
  and\ \bibinfo {author} {\bibfnamefont {Y.~K.}\ \bibnamefont {Kato}},\
  }\bibfield  {title} {\bibinfo {title} {Exciton diffusion, end quenching, and
  exciton-exciton annihilation in individual air-suspended carbon nanotubes},\
  }\href {\doibase 10.1103/physrevb.91.125427} {\bibfield  {journal} {\bibinfo
  {journal} {Phys. Rev. B}\ }\textbf {\bibinfo {volume} {91}},\ \bibinfo
  {pages} {125427} (\bibinfo {year} {2015})}\BibitemShut {NoStop}%
\bibitem [{\citenamefont {Otsuka}\ \emph {et~al.}(2021)\citenamefont {Otsuka},
  \citenamefont {Fang}, \citenamefont {Yamashita}, \citenamefont {Taniguchi},
  \citenamefont {Watanabe},\ and\ \citenamefont {Kato}}]{Otsuka2021}%
  \BibitemOpen
  \bibfield  {author} {\bibinfo {author} {\bibfnamefont {K.}~\bibnamefont
  {Otsuka}}, \bibinfo {author} {\bibfnamefont {N.}~\bibnamefont {Fang}},
  \bibinfo {author} {\bibfnamefont {D.}~\bibnamefont {Yamashita}}, \bibinfo
  {author} {\bibfnamefont {T.}~\bibnamefont {Taniguchi}}, \bibinfo {author}
  {\bibfnamefont {K.}~\bibnamefont {Watanabe}}, \ and\ \bibinfo {author}
  {\bibfnamefont {Y.~K.}\ \bibnamefont {Kato}},\ }\bibfield  {title} {\bibinfo
  {title} {Deterministic transfer of optical-quality carbon nanotubes for
  atomically defined technology},\ }\href {\doibase 10.1038/s41467-021-23413-4}
  {\bibfield  {journal} {\bibinfo  {journal} {Nat. Commun.}\ }\textbf {\bibinfo
  {volume} {12}},\ \bibinfo {pages} {3138} (\bibinfo {year}
  {2021})}\BibitemShut {NoStop}%
\bibitem [{\citenamefont {Puretzky}\ \emph {et~al.}(2005)\citenamefont
  {Puretzky}, \citenamefont {Geohegan}, \citenamefont {Jesse}, \citenamefont
  {Ivanov},\ and\ \citenamefont {Eres}}]{Puretzky2005}%
  \BibitemOpen
  \bibfield  {author} {\bibinfo {author} {\bibfnamefont {A.}~\bibnamefont
  {Puretzky}}, \bibinfo {author} {\bibfnamefont {D.}~\bibnamefont {Geohegan}},
  \bibinfo {author} {\bibfnamefont {S.}~\bibnamefont {Jesse}}, \bibinfo
  {author} {\bibfnamefont {I.}~\bibnamefont {Ivanov}}, \ and\ \bibinfo {author}
  {\bibfnamefont {G.}~\bibnamefont {Eres}},\ }\bibfield  {title} {\bibinfo
  {title} {In situ measurements and modeling of carbon nanotube array growth
  kinetics during chemical vapor deposition},\ }\href {\doibase
  10.1007/s00339-005-3256-7} {\bibfield  {journal} {\bibinfo  {journal} {Appl.
  Phys. A}\ }\textbf {\bibinfo {volume} {81}},\ \bibinfo {pages} {223}
  (\bibinfo {year} {2005})}\BibitemShut {NoStop}%
\bibitem [{\citenamefont {Yoshikawa}\ \emph {et~al.}(2019)\citenamefont
  {Yoshikawa}, \citenamefont {Hisama}, \citenamefont {Ukai}, \citenamefont
  {Takagi}, \citenamefont {Inoue}, \citenamefont {Chiashi},\ and\ \citenamefont
  {Maruyama}}]{Yoshikawa2019}%
  \BibitemOpen
  \bibfield  {author} {\bibinfo {author} {\bibfnamefont {R.}~\bibnamefont
  {Yoshikawa}}, \bibinfo {author} {\bibfnamefont {K.}~\bibnamefont {Hisama}},
  \bibinfo {author} {\bibfnamefont {H.}~\bibnamefont {Ukai}}, \bibinfo {author}
  {\bibfnamefont {Y.}~\bibnamefont {Takagi}}, \bibinfo {author} {\bibfnamefont
  {T.}~\bibnamefont {Inoue}}, \bibinfo {author} {\bibfnamefont
  {S.}~\bibnamefont {Chiashi}}, \ and\ \bibinfo {author} {\bibfnamefont
  {S.}~\bibnamefont {Maruyama}},\ }\bibfield  {title} {\bibinfo {title}
  {Molecular dynamics of chirality definable growth of single-walled carbon
  nanotubes},\ }\href {\doibase 10.1021/acsnano.8b09754} {\bibfield  {journal}
  {\bibinfo  {journal} {ACS Nano}\ }\textbf {\bibinfo {volume} {13}},\ \bibinfo
  {pages} {6506} (\bibinfo {year} {2019})}\BibitemShut {NoStop}%
\bibitem [{\citenamefont {Zhang}\ \emph {et~al.}(2005)\citenamefont {Zhang},
  \citenamefont {Mann}, \citenamefont {Zhang}, \citenamefont {Javey},
  \citenamefont {Li}, \citenamefont {Yenilmez}, \citenamefont {Wang},
  \citenamefont {McVittie}, \citenamefont {Nishi}, \citenamefont {Gibbons},\
  and\ \citenamefont {Dai}}]{Zhang2005}%
  \BibitemOpen
  \bibfield  {author} {\bibinfo {author} {\bibfnamefont {G.}~\bibnamefont
  {Zhang}}, \bibinfo {author} {\bibfnamefont {D.}~\bibnamefont {Mann}},
  \bibinfo {author} {\bibfnamefont {L.}~\bibnamefont {Zhang}}, \bibinfo
  {author} {\bibfnamefont {A.}~\bibnamefont {Javey}}, \bibinfo {author}
  {\bibfnamefont {Y.}~\bibnamefont {Li}}, \bibinfo {author} {\bibfnamefont
  {E.}~\bibnamefont {Yenilmez}}, \bibinfo {author} {\bibfnamefont
  {Q.}~\bibnamefont {Wang}}, \bibinfo {author} {\bibfnamefont {J.~P.}\
  \bibnamefont {McVittie}}, \bibinfo {author} {\bibfnamefont {Y.}~\bibnamefont
  {Nishi}}, \bibinfo {author} {\bibfnamefont {J.}~\bibnamefont {Gibbons}}, \
  and\ \bibinfo {author} {\bibfnamefont {H.}~\bibnamefont {Dai}},\ }\bibfield
  {title} {\bibinfo {title} {Ultra-high-yield growth of vertical single-walled
  carbon nanotubes: Hidden roles of hydrogen and oxygen},\ }\href {\doibase
  10.1073/pnas.0507064102} {\bibfield  {journal} {\bibinfo  {journal} {Proc.
  Natl. Acad. Sci. U.S.A.}\ }\textbf {\bibinfo {volume} {102}},\ \bibinfo
  {pages} {16141} (\bibinfo {year} {2005})}\BibitemShut {NoStop}%
\bibitem [{\citenamefont {Nasibulin}\ \emph {et~al.}(2006)\citenamefont
  {Nasibulin}, \citenamefont {Brown}, \citenamefont {Queipo}, \citenamefont
  {Gonzalez}, \citenamefont {Jiang},\ and\ \citenamefont
  {Kauppinen}}]{Nasibulin2006}%
  \BibitemOpen
  \bibfield  {author} {\bibinfo {author} {\bibfnamefont {A.~G.}\ \bibnamefont
  {Nasibulin}}, \bibinfo {author} {\bibfnamefont {D.~P.}\ \bibnamefont
  {Brown}}, \bibinfo {author} {\bibfnamefont {P.}~\bibnamefont {Queipo}},
  \bibinfo {author} {\bibfnamefont {D.}~\bibnamefont {Gonzalez}}, \bibinfo
  {author} {\bibfnamefont {H.}~\bibnamefont {Jiang}}, \ and\ \bibinfo {author}
  {\bibfnamefont {E.~I.}\ \bibnamefont {Kauppinen}},\ }\bibfield  {title}
  {\bibinfo {title} {An essential role of {CO}$_2$ and {H}$_2${O} during
  single-walled {CNT} synthesis from carbon monoxide},\ }\href {\doibase
  10.1016/j.cplett.2005.10.022} {\bibfield  {journal} {\bibinfo  {journal}
  {Chem. Phys. Lett.}\ }\textbf {\bibinfo {volume} {417}},\ \bibinfo {pages}
  {179} (\bibinfo {year} {2006})}\BibitemShut {NoStop}%
\bibitem [{\citenamefont {Futaba}\ \emph {et~al.}(2009)\citenamefont {Futaba},
  \citenamefont {Goto}, \citenamefont {Yasuda}, \citenamefont {Yamada},
  \citenamefont {Yumura},\ and\ \citenamefont {Hata}}]{Futaba2009}%
  \BibitemOpen
  \bibfield  {author} {\bibinfo {author} {\bibfnamefont {D.~N.}\ \bibnamefont
  {Futaba}}, \bibinfo {author} {\bibfnamefont {J.}~\bibnamefont {Goto}},
  \bibinfo {author} {\bibfnamefont {S.}~\bibnamefont {Yasuda}}, \bibinfo
  {author} {\bibfnamefont {T.}~\bibnamefont {Yamada}}, \bibinfo {author}
  {\bibfnamefont {M.}~\bibnamefont {Yumura}}, \ and\ \bibinfo {author}
  {\bibfnamefont {K.}~\bibnamefont {Hata}},\ }\bibfield  {title} {\bibinfo
  {title} {General rules governing the highly efficient growth of carbon
  nanotubes},\ }\href {\doibase 10.1002/adma.200901257} {\bibfield  {journal}
  {\bibinfo  {journal} {Adv. Mater.}\ }\textbf {\bibinfo {volume} {21}},\
  \bibinfo {pages} {4811} (\bibinfo {year} {2009})}\BibitemShut {NoStop}%
\bibitem [{\citenamefont {Koyano}\ \emph {et~al.}(2019)\citenamefont {Koyano},
  \citenamefont {Inoue}, \citenamefont {Yamamoto}, \citenamefont {Otsuka},
  \citenamefont {Xiang}, \citenamefont {Chiashi},\ and\ \citenamefont
  {Maruyama}}]{Koyano2019}%
  \BibitemOpen
  \bibfield  {author} {\bibinfo {author} {\bibfnamefont {B.}~\bibnamefont
  {Koyano}}, \bibinfo {author} {\bibfnamefont {T.}~\bibnamefont {Inoue}},
  \bibinfo {author} {\bibfnamefont {S.}~\bibnamefont {Yamamoto}}, \bibinfo
  {author} {\bibfnamefont {K.}~\bibnamefont {Otsuka}}, \bibinfo {author}
  {\bibfnamefont {R.}~\bibnamefont {Xiang}}, \bibinfo {author} {\bibfnamefont
  {S.}~\bibnamefont {Chiashi}}, \ and\ \bibinfo {author} {\bibfnamefont
  {S.}~\bibnamefont {Maruyama}},\ }\bibfield  {title} {\bibinfo {title}
  {Regrowth and catalytic etching of individual single-walled carbon nanotubes
  studied by isotope labeling and growth interruption},\ }\href {\doibase
  10.1016/j.carbon.2019.09.031} {\bibfield  {journal} {\bibinfo  {journal}
  {Carbon}\ }\textbf {\bibinfo {volume} {155}},\ \bibinfo {pages} {635}
  (\bibinfo {year} {2019})}\BibitemShut {NoStop}%
\bibitem [{\citenamefont {Xiang}\ \emph {et~al.}(2009)\citenamefont {Xiang},
  \citenamefont {Einarsson}, \citenamefont {Okawa}, \citenamefont {Miyauchi},\
  and\ \citenamefont {Maruyama}}]{Xiang2009}%
  \BibitemOpen
  \bibfield  {author} {\bibinfo {author} {\bibfnamefont {R.}~\bibnamefont
  {Xiang}}, \bibinfo {author} {\bibfnamefont {E.}~\bibnamefont {Einarsson}},
  \bibinfo {author} {\bibfnamefont {J.}~\bibnamefont {Okawa}}, \bibinfo
  {author} {\bibfnamefont {Y.}~\bibnamefont {Miyauchi}}, \ and\ \bibinfo
  {author} {\bibfnamefont {S.}~\bibnamefont {Maruyama}},\ }\bibfield  {title}
  {\bibinfo {title} {Acetylene-accelerated alcohol catalytic chemical vapor
  deposition growth of vertically aligned single-walled carbon nanotubes},\
  }\href {\doibase 10.1021/jp810454f} {\bibfield  {journal} {\bibinfo
  {journal} {J. Phys. Chem. C}\ }\textbf {\bibinfo {volume} {113}},\ \bibinfo
  {pages} {7511} (\bibinfo {year} {2009})}\BibitemShut {NoStop}%
\bibitem [{\citenamefont {Ding}\ \emph
  {et~al.}(2009{\natexlab{b}})\citenamefont {Ding}, \citenamefont {Tselev},
  \citenamefont {Wang}, \citenamefont {Yuan}, \citenamefont {Chu},
  \citenamefont {McNicholas}, \citenamefont {Li},\ and\ \citenamefont
  {Liu}}]{Ding2009a}%
  \BibitemOpen
  \bibfield  {author} {\bibinfo {author} {\bibfnamefont {L.}~\bibnamefont
  {Ding}}, \bibinfo {author} {\bibfnamefont {A.}~\bibnamefont {Tselev}},
  \bibinfo {author} {\bibfnamefont {J.}~\bibnamefont {Wang}}, \bibinfo {author}
  {\bibfnamefont {D.}~\bibnamefont {Yuan}}, \bibinfo {author} {\bibfnamefont
  {H.}~\bibnamefont {Chu}}, \bibinfo {author} {\bibfnamefont {T.~P.}\
  \bibnamefont {McNicholas}}, \bibinfo {author} {\bibfnamefont
  {Y.}~\bibnamefont {Li}}, \ and\ \bibinfo {author} {\bibfnamefont
  {J.}~\bibnamefont {Liu}},\ }\bibfield  {title} {\bibinfo {title} {Selective
  growth of well-aligned semiconducting single-walled carbon nanotubes},\
  }\href {\doibase 10.1021/nl803496s} {\bibfield  {journal} {\bibinfo
  {journal} {Nano Lett.}\ }\textbf {\bibinfo {volume} {9}},\ \bibinfo {pages}
  {800} (\bibinfo {year} {2009}{\natexlab{b}})}\BibitemShut {NoStop}%
\bibitem [{\citenamefont {Kimura}\ \emph {et~al.}(2018)\citenamefont {Kimura},
  \citenamefont {Hijikata}, \citenamefont {Eveleens}, \citenamefont {Page},\
  and\ \citenamefont {Irle}}]{Kimura2018}%
  \BibitemOpen
  \bibfield  {author} {\bibinfo {author} {\bibfnamefont {R.}~\bibnamefont
  {Kimura}}, \bibinfo {author} {\bibfnamefont {Y.}~\bibnamefont {Hijikata}},
  \bibinfo {author} {\bibfnamefont {C.~A.}\ \bibnamefont {Eveleens}}, \bibinfo
  {author} {\bibfnamefont {A.~J.}\ \bibnamefont {Page}}, \ and\ \bibinfo
  {author} {\bibfnamefont {S.}~\bibnamefont {Irle}},\ }\bibfield  {title}
  {\bibinfo {title} {Chiral-selective etching effects on carbon nanotube growth
  at edge carbon atoms},\ }\href {\doibase 10.1002/jcc.25610} {\bibfield
  {journal} {\bibinfo  {journal} {J. Comput. Chem.}\ }\textbf {\bibinfo
  {volume} {40}},\ \bibinfo {pages} {375} (\bibinfo {year} {2018})}\BibitemShut
  {NoStop}%
\bibitem [{\citenamefont {Wang}\ \emph {et~al.}(2018)\citenamefont {Wang},
  \citenamefont {Jin}, \citenamefont {Liu}, \citenamefont {Yu}, \citenamefont
  {Ji}, \citenamefont {Wei}, \citenamefont {Zhang}, \citenamefont {Zhang},
  \citenamefont {Li}, \citenamefont {Yuan}, \citenamefont {Li}, \citenamefont
  {Liu}, \citenamefont {Wu}, \citenamefont {Wei}, \citenamefont {Wang},
  \citenamefont {Li}, \citenamefont {Zhang}, \citenamefont {Kong},
  \citenamefont {Fan},\ and\ \citenamefont {Jiang}}]{Wang2018}%
  \BibitemOpen
  \bibfield  {author} {\bibinfo {author} {\bibfnamefont {J.}~\bibnamefont
  {Wang}}, \bibinfo {author} {\bibfnamefont {X.}~\bibnamefont {Jin}}, \bibinfo
  {author} {\bibfnamefont {Z.}~\bibnamefont {Liu}}, \bibinfo {author}
  {\bibfnamefont {G.}~\bibnamefont {Yu}}, \bibinfo {author} {\bibfnamefont
  {Q.}~\bibnamefont {Ji}}, \bibinfo {author} {\bibfnamefont {H.}~\bibnamefont
  {Wei}}, \bibinfo {author} {\bibfnamefont {J.}~\bibnamefont {Zhang}}, \bibinfo
  {author} {\bibfnamefont {K.}~\bibnamefont {Zhang}}, \bibinfo {author}
  {\bibfnamefont {D.}~\bibnamefont {Li}}, \bibinfo {author} {\bibfnamefont
  {Z.}~\bibnamefont {Yuan}}, \bibinfo {author} {\bibfnamefont {J.}~\bibnamefont
  {Li}}, \bibinfo {author} {\bibfnamefont {P.}~\bibnamefont {Liu}}, \bibinfo
  {author} {\bibfnamefont {Y.}~\bibnamefont {Wu}}, \bibinfo {author}
  {\bibfnamefont {Y.}~\bibnamefont {Wei}}, \bibinfo {author} {\bibfnamefont
  {J.}~\bibnamefont {Wang}}, \bibinfo {author} {\bibfnamefont {Q.}~\bibnamefont
  {Li}}, \bibinfo {author} {\bibfnamefont {L.}~\bibnamefont {Zhang}}, \bibinfo
  {author} {\bibfnamefont {J.}~\bibnamefont {Kong}}, \bibinfo {author}
  {\bibfnamefont {S.}~\bibnamefont {Fan}}, \ and\ \bibinfo {author}
  {\bibfnamefont {K.}~\bibnamefont {Jiang}},\ }\bibfield  {title} {\bibinfo
  {title} {Growing highly pure semiconducting carbon nanotubes by
  electrotwisting the helicity},\ }\href {\doibase 10.1038/s41929-018-0057-x}
  {\bibfield  {journal} {\bibinfo  {journal} {Nat. Catal.}\ }\textbf {\bibinfo
  {volume} {1}},\ \bibinfo {pages} {326} (\bibinfo {year} {2018})}\BibitemShut
  {NoStop}%
\bibitem [{\citenamefont {Hussain}\ \emph {et~al.}(2018)\citenamefont
  {Hussain}, \citenamefont {Liao}, \citenamefont {Zhang}, \citenamefont {Ding},
  \citenamefont {Laiho}, \citenamefont {Ahmad}, \citenamefont {Wei},
  \citenamefont {Tian}, \citenamefont {Jiang},\ and\ \citenamefont
  {Kauppinen}}]{Hussain2018}%
  \BibitemOpen
  \bibfield  {author} {\bibinfo {author} {\bibfnamefont {A.}~\bibnamefont
  {Hussain}}, \bibinfo {author} {\bibfnamefont {Y.}~\bibnamefont {Liao}},
  \bibinfo {author} {\bibfnamefont {Q.}~\bibnamefont {Zhang}}, \bibinfo
  {author} {\bibfnamefont {E.-X.}\ \bibnamefont {Ding}}, \bibinfo {author}
  {\bibfnamefont {P.}~\bibnamefont {Laiho}}, \bibinfo {author} {\bibfnamefont
  {S.}~\bibnamefont {Ahmad}}, \bibinfo {author} {\bibfnamefont
  {N.}~\bibnamefont {Wei}}, \bibinfo {author} {\bibfnamefont {Y.}~\bibnamefont
  {Tian}}, \bibinfo {author} {\bibfnamefont {H.}~\bibnamefont {Jiang}}, \ and\
  \bibinfo {author} {\bibfnamefont {E.~I.}\ \bibnamefont {Kauppinen}},\
  }\bibfield  {title} {\bibinfo {title} {Floating catalyst {CVD} synthesis of
  single walled carbon nanotubes from ethylene for high performance transparent
  electrodes},\ }\href {\doibase 10.1039/c8nr00716k} {\bibfield  {journal}
  {\bibinfo  {journal} {Nanoscale}\ }\textbf {\bibinfo {volume} {10}},\
  \bibinfo {pages} {9752} (\bibinfo {year} {2018})}\BibitemShut {NoStop}%
\bibitem [{\citenamefont {He}\ \emph {et~al.}(2018)\citenamefont {He},
  \citenamefont {Magnin}, \citenamefont {Jiang}, \citenamefont {Amara},
  \citenamefont {Kauppinen}, \citenamefont {Loiseau},\ and\ \citenamefont
  {Bichara}}]{He2018}%
  \BibitemOpen
  \bibfield  {author} {\bibinfo {author} {\bibfnamefont {M.}~\bibnamefont
  {He}}, \bibinfo {author} {\bibfnamefont {Y.}~\bibnamefont {Magnin}}, \bibinfo
  {author} {\bibfnamefont {H.}~\bibnamefont {Jiang}}, \bibinfo {author}
  {\bibfnamefont {H.}~\bibnamefont {Amara}}, \bibinfo {author} {\bibfnamefont
  {E.~I.}\ \bibnamefont {Kauppinen}}, \bibinfo {author} {\bibfnamefont
  {A.}~\bibnamefont {Loiseau}}, \ and\ \bibinfo {author} {\bibfnamefont
  {C.}~\bibnamefont {Bichara}},\ }\bibfield  {title} {\bibinfo {title} {Growth
  modes and chiral selectivity of single-walled carbon nanotubes},\ }\href
  {\doibase 10.1039/c7nr09539b} {\bibfield  {journal} {\bibinfo  {journal}
  {Nanoscale}\ }\textbf {\bibinfo {volume} {10}},\ \bibinfo {pages} {6744}
  (\bibinfo {year} {2018})}\BibitemShut {NoStop}%
\bibitem [{\citenamefont {Takagi}\ \emph {et~al.}(2006)\citenamefont {Takagi},
  \citenamefont {Homma}, \citenamefont {Suzuki},\ and\ \citenamefont
  {Kobayashi}}]{Takagi2006}%
  \BibitemOpen
  \bibfield  {author} {\bibinfo {author} {\bibfnamefont {D.}~\bibnamefont
  {Takagi}}, \bibinfo {author} {\bibfnamefont {Y.}~\bibnamefont {Homma}},
  \bibinfo {author} {\bibfnamefont {S.}~\bibnamefont {Suzuki}}, \ and\ \bibinfo
  {author} {\bibfnamefont {Y.}~\bibnamefont {Kobayashi}},\ }\bibfield  {title}
  {\bibinfo {title} {In situ scanning electron microscopy of single-walled
  carbon nanotube growth},\ }\href {\doibase 10.1002/sia.2448} {\bibfield
  {journal} {\bibinfo  {journal} {Surf. Interface Anal.}\ }\textbf {\bibinfo
  {volume} {38}},\ \bibinfo {pages} {1743} (\bibinfo {year}
  {2006})}\BibitemShut {NoStop}%
\bibitem [{\citenamefont {Inoue}\ \emph {et~al.}(2013)\citenamefont {Inoue},
  \citenamefont {Hasegawa}, \citenamefont {Badar}, \citenamefont {Aikawa},
  \citenamefont {Chiashi},\ and\ \citenamefont {Maruyama}}]{Inoue2013}%
  \BibitemOpen
  \bibfield  {author} {\bibinfo {author} {\bibfnamefont {T.}~\bibnamefont
  {Inoue}}, \bibinfo {author} {\bibfnamefont {D.}~\bibnamefont {Hasegawa}},
  \bibinfo {author} {\bibfnamefont {S.}~\bibnamefont {Badar}}, \bibinfo
  {author} {\bibfnamefont {S.}~\bibnamefont {Aikawa}}, \bibinfo {author}
  {\bibfnamefont {S.}~\bibnamefont {Chiashi}}, \ and\ \bibinfo {author}
  {\bibfnamefont {S.}~\bibnamefont {Maruyama}},\ }\bibfield  {title} {\bibinfo
  {title} {Effect of gas pressure on the density of horizontally aligned
  single-walled carbon nanotubes grown on quartz substrates},\ }\href {\doibase
  10.1021/jp401681e} {\bibfield  {journal} {\bibinfo  {journal} {J. Phys. Chem.
  C}\ }\textbf {\bibinfo {volume} {117}},\ \bibinfo {pages} {11804} (\bibinfo
  {year} {2013})}\BibitemShut {NoStop}%
\bibitem [{\citenamefont {Einarsson}\ \emph {et~al.}(2008)\citenamefont
  {Einarsson}, \citenamefont {Murakami}, \citenamefont {Kadowaki},\ and\
  \citenamefont {Maruyama}}]{Einarsson2008}%
  \BibitemOpen
  \bibfield  {author} {\bibinfo {author} {\bibfnamefont {E.}~\bibnamefont
  {Einarsson}}, \bibinfo {author} {\bibfnamefont {Y.}~\bibnamefont {Murakami}},
  \bibinfo {author} {\bibfnamefont {M.}~\bibnamefont {Kadowaki}}, \ and\
  \bibinfo {author} {\bibfnamefont {S.}~\bibnamefont {Maruyama}},\ }\bibfield
  {title} {\bibinfo {title} {Growth dynamics of vertically aligned
  single-walled carbon nanotubes from in situ measurements},\ }\href {\doibase
  10.1016/j.carbon.2008.02.021} {\bibfield  {journal} {\bibinfo  {journal}
  {Carbon}\ }\textbf {\bibinfo {volume} {46}},\ \bibinfo {pages} {923}
  (\bibinfo {year} {2008})}\BibitemShut {NoStop}%
\bibitem [{\citenamefont {Bachilo}\ \emph {et~al.}(2002)\citenamefont
  {Bachilo}, \citenamefont {Strano}, \citenamefont {Kittrell}, \citenamefont
  {Hauge}, \citenamefont {Smalley},\ and\ \citenamefont
  {Weisman}}]{Bachilo2002}%
  \BibitemOpen
  \bibfield  {author} {\bibinfo {author} {\bibfnamefont {S.~M.}\ \bibnamefont
  {Bachilo}}, \bibinfo {author} {\bibfnamefont {M.~S.}\ \bibnamefont {Strano}},
  \bibinfo {author} {\bibfnamefont {C.}~\bibnamefont {Kittrell}}, \bibinfo
  {author} {\bibfnamefont {R.~H.}\ \bibnamefont {Hauge}}, \bibinfo {author}
  {\bibfnamefont {R.~E.}\ \bibnamefont {Smalley}}, \ and\ \bibinfo {author}
  {\bibfnamefont {R.~B.}\ \bibnamefont {Weisman}},\ }\bibfield  {title}
  {\bibinfo {title} {Structure-assigned optical spectra of single-walled carbon
  nanotubes},\ }\href {\doibase 10.1126/science.1078727} {\bibfield  {journal}
  {\bibinfo  {journal} {Science}\ }\textbf {\bibinfo {volume} {298}},\ \bibinfo
  {pages} {2361} (\bibinfo {year} {2002})}\BibitemShut {NoStop}%
\bibitem [{\citenamefont {Liao}\ \emph {et~al.}(2018)\citenamefont {Liao},
  \citenamefont {Jiang}, \citenamefont {Wei}, \citenamefont {Laiho},
  \citenamefont {Zhang}, \citenamefont {Khan},\ and\ \citenamefont
  {Kauppinen}}]{Liao2018}%
  \BibitemOpen
  \bibfield  {author} {\bibinfo {author} {\bibfnamefont {Y.}~\bibnamefont
  {Liao}}, \bibinfo {author} {\bibfnamefont {H.}~\bibnamefont {Jiang}},
  \bibinfo {author} {\bibfnamefont {N.}~\bibnamefont {Wei}}, \bibinfo {author}
  {\bibfnamefont {P.}~\bibnamefont {Laiho}}, \bibinfo {author} {\bibfnamefont
  {Q.}~\bibnamefont {Zhang}}, \bibinfo {author} {\bibfnamefont {S.~A.}\
  \bibnamefont {Khan}}, \ and\ \bibinfo {author} {\bibfnamefont {E.~I.}\
  \bibnamefont {Kauppinen}},\ }\bibfield  {title} {\bibinfo {title} {Direct
  synthesis of colorful single-walled carbon nanotube thin films},\ }\href
  {\doibase 10.1021/jacs.8b05151} {\bibfield  {journal} {\bibinfo  {journal}
  {J. Am. Chem. Soc.}\ }\textbf {\bibinfo {volume} {140}},\ \bibinfo {pages}
  {9797} (\bibinfo {year} {2018})}\BibitemShut {NoStop}%
\end{thebibliography}
\end{document}